\documentclass[preprint]{aastex}

\slugcomment{Submitted to ApJ.}
\shorttitle{Kerr Black Hole Tori}
\shortauthors{De Villiers and Hawley}

\begin{document}

\title{Three-Dimensional Hydrodynamic Simulations of Accretion
Tori in Kerr Spacetimes}


\author{Jean-Pierre De Villiers and John F. Hawley}
\affil{Astronomy Department\\
University of Virginia\\ 
P.O. Box 3818, University Station\\
Charlottesville, VA 22903-0818}
\email{jd5v@virginia.edu; jh8h@virginia.edu}

\begin{abstract} 
This paper presents results of three-dimensional simulations of
global hydrodynamic instabilities in black hole tori, extending earlier work by
Hawley to Kerr spacetimes.  This study probes a three-dimensional parameter
space of torus angular momentum, torus size, and black hole angular momentum.
We have observed the growth of the Papaloizou-Pringle instability for a range of
torus configurations and the resultant formation of m=1 planets.  We have also
observed the quenching of this instability in the presence of early accretion
flows; however, in one simulation both early accretion and planet formation
occurred.  Though most of the conclusions reached in Hawley's earlier work on
Schwarzschild black holes carry over to Kerr spacetime, the presence of frame
dragging in the Kerr geometry adds an element of complexity to the simulations;
we have seen especially clear examples of this phenomenon in the accretion flows
that arise from retrograde tori.  
\end{abstract}



\section{Introduction}

More than three decades have passed since the first suggestions that black hole
accretion is a ubiquitous astrophysical phenomenon (e.g.  Lynden-Bell 1969), and
there is now substantial and compelling observational evidence for it.  The
range of permissible masses for compact, unseen companions puts many binary
systems squarely in the black hole domain (e.g.  Cyg X-1, A0620-00; van Paradijs
\& McClintock 1995), while stellar kinematical observations of galactic cores
(reviewed by Tremaine 1997) imply large amounts of mass ($10^6-10^9\, M_\odot$)
concentrated in small (parsec-sized) volumes.  Indirect indications of accretion
include copious nonthermal radiation from active galactic nuclei, relativistic
expansion, and jets.  Iron line observations (e.g., Tanaka et al.  1995) show
how, in principle, observations can probe directly the physics of the
strong-field region, distinguishing black hole accretion from accretion onto
white dwarfs or neutron stars.

With the emergence of a unified model to account for the wide variety
of observational characteristics of both X-ray binaries and active
galactic nuclei (AGN), it becomes increasingly important to support
theory and observation with detailed numerical simulations of the
physics of black hole accretion.  Distinctive and astrophysically
interesting effects are expected from accretion flows in the Kerr
metric, and modeling will require development of new fully
general relativistic (GR) simulation codes.
This becomes especially relevant in light of results
that suggest that most galactic core black holes are rotating
(Elvis, Risaliti, \& Zamorani 2002).

Most of the simulations to date of relativistic black hole accretion
have been for hydrodynamics alone.  Extensive work was done in the
1980s on the numerical simulation of hydrodynamics around black holes
(Hawley, Smarr \& Wilson 1984a, 1984b; hereafter HSWa and HSWb).  This
work was based on numerical techniques developed through the pioneering
efforts of Wilson (1972).  The HSWb GR
hydrodynamics code was written in spherical polar  Boyer-Lindquist
coordinates, and used operator-split finite-differencing on a staggered
grid, with piecewise linear representations for the fundamental
variables.  The GR hydrodynamic studies using this code (HSWb; Hawley
\& Smarr 1986; Hawley 1986) were two-dimensional (assuming axisymmetry)
and dealt with the prompt infall of rotating gas toward a central black
hole, the role of the centrifugal barrier, centrifugally-driven
outflows, and the formation of pressure supported thick disks or tori.
Since then, there have been other axisymmetric GR hydrodynamic
simulations using a similar numerical approach (Yokosawa 1995;
Igumenshchev \& Beloborodov 1997).

It is now recognized that magnetic fields play an essential role in
the outward transport of angular momentum in accretion disks through
the action of the magnetorotational instability of Balbus and Hawley
(1991).  This implies the need for a general relativistic
magnetohydrodynamics (MHD) simulation code.  Here again, Wilson did
pioneering work, carrying out two-dimensional simulations of magnetized
accretion over two and a half decades ago (Wilson 1975; 1977; 1978).
The task was arguably beyond the computers then available, however.
Only recently have research groups returned to full GR MHD
simulations (e.g., Koide, Shibata, \& Kudoh 1999; Komissarov 1999).

Rather than employing full relativity, the most recent
three-dimensional global black hole MHD accretion simulations (e.g.,
Hawley \& Balbus 2002) use a pseudo-Newtonian potential that can
emulate certain important characteristics of the Schwarzschild metric.
Computing a three dimensional MHD accretion flow in a full Kerr metric
is a more ambitious undertaking.  It is our aim to develop a new
three-dimensional GR MHD code to enable such global accretion disk
studies.  This paper represents a first step, specifically the initial
development and application of the hydrodynamic portion of a Kerr
metric GR code.

The culmination of the earlier GR hydrodynamics effort was a fully
three-dimensional numerical investigation (Hawley 1991; hereafter H91)
of the Papaloizou-Pringle instability (Papaploizou \& Pringle 1984) for
accretion tori in a Schwarzschild metric.  In this paper we extend
those simulations to Kerr (rotating) black holes, and consider the
effect of black hole angular momentum on prograde and retrograde torus
orbits.  The Papaloizou-Pringle instability (PPI) remains an
interesting topic of research, and it provides a nontrivial application
of full three-dimensional general-relativistic hydrodynamics.  

The plan of this paper is as follows.  In \S2 we outline the equations
and the general numerical procedures.  In \S3 we discuss the properties
of equilibrium gas tori in the Kerr metric, and the properties of the
Papaloizou-Pringle instability to which these tori are vulnerable.  The
results of a series of numerical simulations of tori in the Kerr metric
are presented in \S4, and these results are discussed in \S5.

\section{Equations and Numerical Methods}

We wish to study the evolution of a fluid in the background spacetime
of a Kerr (rotating) black hole.   We adopt Boyer-Lindquist 
coordinates, $(t,r,\theta,\phi)$,  
for which the line element has the form,
\begin{equation}\label{kerr}
{ds}^2=g_{t t}\,{dt}^2+2\,g_{t \phi}\,{dt}\,{d \phi}+g_{r r}\,{dr}^2+
 g_{\theta \theta}\,{d \theta}^2+g_{\phi \phi}\,{d \phi}^2 .
\end{equation}
In keeping with Misner, Thorne, \& Wheeler (1973), we use the metric signature
$(-,+,+,+)$, along with geometrodynamic units where $G = c = 1$; the black hole
mass is unity, $M=1$.  The determinant of the 4-metric is $g$, and $\sqrt{-g} =
\alpha\,\sqrt{\gamma}$ where $\alpha$ is the lapse function,
$\alpha=1/\sqrt{-g^{tt}}$, and $\gamma$ is the determinant of the spatial
$3$-metric.

For a relativistic test fluid described by a density $\rho$, specific internal
energy $\epsilon$, and $4$-velocity $U^\mu$, 
we define the transport velocity $V^\mu$ 
as follows:
\begin{equation}\label{tranvel}
 V^\mu = {U^\mu \over U^t} ,
\end{equation}
where $U^t = W/\alpha$, and $W$ is the gravitational redshift factor,
\begin{equation}\label{gred}
 W = {1 \over (1-V^\mu\,V_\mu)^{1/2}} .
\end{equation}
We also define the momentum,
\begin{equation}\label{momdef}
 S_\mu = \rho\,h\,W\,U_\mu ,
\end{equation}
and auxiliary density and energy functions
$D = \rho\,W$ and $E = D\,\epsilon$. 

Using these variables the conservation laws can be rewritten into
a form suitable for finite differencing.
The equation for mass conservation is written
\begin{equation}\label{masscons}
\partial_t\,D + {1 \over
\sqrt{\gamma}}\,\partial_j\,(D\,\sqrt{\gamma}\,V^j)  = 0.
\end{equation}
Conservation of the fluid energy-momentum tensor 
$\nabla_\mu {T}^{\mu \nu} = 0$ yields 
momentum conservation equations,
\begin{equation}\label{momcons}
\partial_t\,S_i + {1 \over \sqrt{\gamma}}\,
\partial_j\,(S_i\,\sqrt{\gamma}\,V^j)
+ \alpha\,\partial_j\,P + 
{1 \over 2}\,{S_\mu\,S_\nu \over S^t}\partial_j\,g^{\mu\,\nu}
=  0,
\end{equation}
and an internal energy conservation equation
\begin{equation}\label{encons}
\partial_t\,E + {1 \over
\sqrt{\gamma}}\,\partial_j\,(E\,\sqrt{\gamma}\,V^j) + P\,\partial_t\,W
+ {P \over \sqrt{\gamma}}\,\partial_j\,(W\,\sqrt{\gamma}\,V^j) =  0.
\end{equation}
Spatial indices are indicated by roman characters $i,j=1,2,3$.  
We assume an
ideal gas equation of state $P=\rho\,\epsilon\,(\Gamma-1)$, where
$\Gamma$ is the adiabatic exponent. 
This is the same system of equations as described in HSWa.

The GR hydrodynamics code evolves time-explicit, operator-split, finite
difference forms of equations (\ref{masscons})---(\ref{encons}).  The algorithm
is a three-dimensional generalization of the solver described in HSWb, with some
additional modifications.  For example, velocity renormalization, $U^\mu\,U_\mu
= -1$, which is invoked in the source and transport steps following the solution
of the discretized momentum equation (\ref{momcons}) is now implemented using
the algebraically equivalent condition $S^\mu\,S_\mu = -{(D + \Gamma\,E)}^2$ for
improved numerical stability.  Several improvements are possible because of the
increase in computer power since 1984.  Analytically determined metric terms and
metric derivatives are now calculated and stored at all needed grid locations
rather than averaged.  Array operations have also been streamlined wherever
possible using Fortran 90 array syntax.  The code uses message passing
parallelism with a form of domain decomposition, where the global grid is
partitioned into subgrids, with each subgrid assigned to a processor.  Data on
each subgrid is evolved independently during the source and transport phases of
a timestep and data on subgrid boundaries is exchanged at the end of each phase
through message-passing calls.  This results in a highly scalable code that
exhibits good speedup over the full range of practically realizable subgrids.

The present simulations have been designed as a follow-on to the
results in H91, but unlike the code used there, we do not need to
impose equatorial symmetry.  The calculations performed on a $64 \times
32 \times 64$ $(r,\theta,\phi)$-grid in H91 correspond to a $64 \times
64 \times 64$ grid here.
The ergosphere can be part of the computational domain, but  
at the event horizon, $r_H = M + \sqrt{M^2-a^2}$, some of the
Boyer-Lindquist metric terms are singular, so the inner edge of the
radial grid must lie at some point outside the horizon, continuing out
to some selected outer grid boundary. 
Although the horizon itself can never be part of
the computational grid, tests indicate that it is possible to come
arbitrarily close to the horizon (e.g., $r_{in} = 2.001\,M$ for the
Schwarzschild case).  But this comes at a steep price:  the narrow
region where the metric terms diverge rapidly must be covered by a
large number of grid zones (in the extreme Kerr limit, the number of
zones inside the ergosphere could be as high as the number of zones
that lie outside it).  In addition to consuming large amounts of memory
in a three dimensional simulation, 
these innermost grid zones can significantly
reduce the time step size, greatly increasing the total
computational time.  The specific location of the
inner edge of the grid is determined by a balance between the 
physical phenomena of interest in a simulation and
the memory and performance demands.  For the hydrodynamic torus
simulations considered here, the region near the horizon is of
secondary interest, so the inner grid boundary, $r_{min}$, was set at
or just inside the static limit on the equator, $r_{min} \le
r_{static}{|}_{\theta=\pi/2}=2\,M$.  In the runs reported here, this
ranged from $r_{in} = 1.90\,M$ for the smaller grids to $r_{in} =
2.05\,M$ for the larger grids.
The polar grid
extends almost to the rotational axes at $\theta=0$ and $\pi$, but the axes
themselves are not included.  As in non-relativistic codes, the polar axes are
problematic due to the shrinking of zone volumes, and the collapse of the inner
radial zone face at $r = 0$.  These problems are further compounded by the fact
that, here again, some metric components are singular.  Torus stability studies
do not require the axes, and their omission is not important to the present
simulations.  Regularized operators will be introduced in future work.  The
radial and polar grids use logarithmic scaling for all runs in order to
concentrate zones near the horizon and the equator.  The azimuthal angle $\phi$
spans the full range from $0$ to $2\pi$ with equally-spaced grid zones and
periodic boundary conditions.  Previous studies of the Papaloizou-Pringle
instability show that using the full $\phi$ range {\it is} important as the most
important unstable modes have azimuthal wavenumber $m=1$.

The time step $\Delta t$ is determined by the extremal light-crossing
time for a grid zone, as described in HSWb.  This time step size
remains fixed for the duration of the simulation since it is a purely
geometric result.

The initial state for a simulation is generated from the thick disk
equilibrium solution (described in \S3) with given input parameters.
This initial configuration is then perturbed with small, random
enthalpy fluctuations with a maximum amplitude of 1\%.  These
perturbations are the seeds from which the full PPI will develop.

\section{Tori and the Kerr Metric}

The initial stationary, axisymmetric analytic torus solution is presented in
HSWa.  For convenience, the essential results are repeated here.

The initial state for the simulations is a stationary, axisymmetric solution
($\partial_t = \partial_\phi = 0$) to equations
(\ref{masscons})---(\ref{encons}) with no internal poloidal motion, i.e.
$U^r=U^\theta=0$.  The combined centrifugal and gravitational accelerations
(which together make up an effective potential) are balanced by pressure
gradients, keeping the disk in equilibrium.  To develop this solution, define
the specific angular momentum ($l$) and angular velocity ($\Omega$), as
\begin{equation}
U_\mu = U_t\,\left(1,0,0,-l\right)
\end{equation}
\begin{equation}
U^\mu = U^t\,\left(1,0,0,\Omega\right).
\end{equation}
Applying the orthogonality condition $U^\mu\,U_\mu = -1$ leads directly
to the following expression for $U_t$
\begin{equation}
U_t = -{\left( \| g^{t t} - 2\,l\,g^{t \phi} 
+ l^2\,g^{\phi \phi}\|\right)}^{-1/2}
\end{equation}
which is related to the binding energy, $e_{bind} = -U_t$.
Here, we note that equation (94a) of HSWa should be extended to allow a
more general disk outer boundary, which is specified by the binding
energy at the surface $e_{surf} = -U_{t_{lim}}$.  The more general form
of equation (94a) is
\begin{equation} \epsilon =  {1 \over
\Gamma}\,\left({U_{t_{lim}} \over U_t} - 1\right).  
\end{equation} 
For a constant entropy adiabatic gas the pressure is given by $P =
\rho\,\epsilon\,(\Gamma - 1) = K\,\rho^\Gamma$, and density is given by
$\rho={\left[{\epsilon\,(\Gamma - 1) / K}\right]}^{1/(\Gamma - 1)}$.
These relations completely specify the initial equilibrium torus.

A particular constant angular momentum thick disk solution is specified
by choosing the angular momentum $l$, the binding energy at the surface
of the torus (as above),  and the entropy parameter $K$. 
In keeping with HSWb, $K=0.01$ is fixed for all simulations, so only
the specific angular momentum $l$, surface binding energy $e_{surf}$,
and the Kerr parameter $a$ are varied.

\subsection{The Papaloizou-Pringle Instability}

The stability of tori with constant specific angular momentum $l$ has
been of interest since the work of Papaloizou and Pringle
(1984), who established that these tori are unstable to
non-axisymmetric global modes.  It has since been demonstrated that
although other angular momentum distributions are also unstable, the
constant-$l$ tori are the most susceptible to the PPI.
Subsequent work, notably Narayan, Goldreich, and Goodman (1987),
consolidated the central features of the instability into the following
picture.  The global unstable modes have a co-rotation radius within
the torus; the co-rotation is located in a narrow region where waves
cannot propagate; this region separates inner and outer regions where
wave propagation is possible; waves can tunnel through the corotation
zone and interact with waves in the other region; and the transmitted
modes are amplified only if there is a feedback mechanism, usually in
the form of a reflecting boundary at the inner and/or outer edge of the
torus.

However, linear stability analysis can go only so far, and numerical work is
required to probe the non-linear effects that help determine the final amplitude
of the global modes, and hence the cumulative effect of the instability on the
torus.  Hawley (1987) carried out a relativistic numerical study of the
evolution of the PPI into the fully non-linear regime in 2D height-integrated
disks.  Although this paper verified that the azimuthally averaged rotation law
($\Omega \sim r^{-q}$) exhibits the expected effects of angular momentum
redistribution ($q_{sat} \sim \sqrt{3}$), it also showed that the distribution
of matter was not azimuthally symmetric, but instead took the form of
counter-rotating epicyclic vortices, or ``planets'', with $m$ planets emerging
from the growth of a mode of order $m$.  The stability of thick tori in general
relativity was studied by Blaes and Hawley (1988) who evolved unstable tori in a
two-dimensional $(r,\phi)$ Schwarzshild metric.  This work led to the full three
dimensional simulations in H91.  The present work extends these calculations to
include the effects of nonzero black hole angular momentum.  Although we do not
repeat the linear analysis of Blaes and Hawley (1988) with a non-zero Kerr
parameter $a$, we know from previous work that mode growth depends very
sensitively on the initial torus parameters.  Black hole rotation simply adds
additional complexity.

A point of particular interest has been the effect of  accretion on
the evolution of the PPI.  Blaes (1987) computed growth rates for
two-dimensional tori in the Schwarzschild metric and found that they
went to zero for models with even a small net accretion flow past the
location of the marginally-stable orbit.  It was argued that the loss
of the inner reflecting boundary through the development of an
accretion flow suppresses mode growth.  This result was subsequently
confirmed analytically by Gat and Livio (1992) for a Newtonian
potential.  In the three-dimensional case, things are not so clear
cut.  H91 argues that since the accretion flow is confined to the
equatorial plane, there is still a reflecting boundary above and below
the plane.  However, Dwarkadas \& Balbus (1996) argued that the
stabilizing effect of an accretion flow is due more to the dynamics at
the corotation point than to the absence of reflection.

\subsection{Disk Properties and the Kerr Parameter}

One of the goals of H91 was to study the effect of the proximity of the cusp in
the relativistic potential to the inner edge of the torus, and how accretion
into the black hole affects mode growth.  Thus the majority of the H91 runs
dealt with tori with angular momentum above the marginally bound value, that is
the smallest specific angular momentum for which a particle orbit has an inner
turning point.  In H91, an initial torus configuration was specified by choosing
the angular momentum and the surface binding energy.  Here, the structural
properties of the initial torus are also influenced by the Kerr parameter.  We
therefore need to understand the interdependence of $l$, $U_{t_{lim}}$, and $a$
in order to specify initial configurations for the simulations.

\begin{figure}[ht]
     \epsscale{0.5}
     \plotone{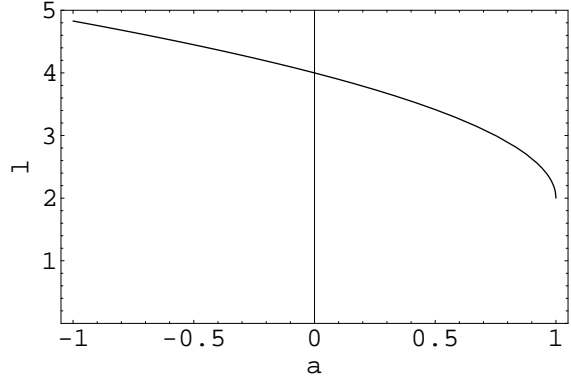}
     \caption{\label{lmb} 
     Plot of $l_{mb}(a)$, the angular momentum of a marginally bound orbit 
     as a function of black hole angular momentum.} 
\end{figure}

To help guide the choice of $l$, it is useful to consider the quantity $l_{mb}$
that gives the specific angular momentum for a test particle on a marginally
bound orbit.
Particles on retrograde orbits require larger absolute values of
the specific angular momentum to remain out of the hole than do particles on
prograde orbits.  This can be seen in Figure \ref{lmb}, which plots the
dependence of $l_{mb}$ on the Kerr parameter $a$; the relation $l_{mb}(a)$ can
be found, for instance, in Frolov and Novikov (1998).  The quantity $l_{mb}$
also helps distinguish families of tori that have stable orbits from those
that do not.  Tori with $l = l_{mb}$ are referred to as marginal tori from
hereon. 

\begin{figure}[ht] 
     \epsscale{1.0}
     \plotone{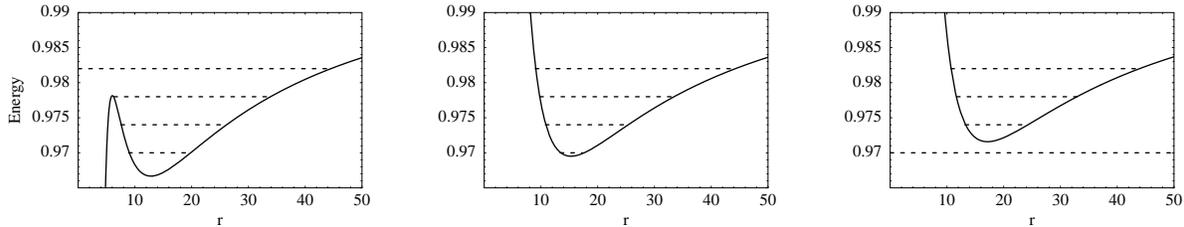} 
     \caption{\label{utlim} Plots of energy function $e = -U_t$ for $l=4.5$. 
     The choice of $-U_{t_{lim}}$ determines the width of the
     disk.  Here, the relevant portion of the energy surface (in the equatorial
     plane) is shown for $a =-0.8$ (left),  $a =0.0$ (middle), $a =1.0$ (right),
     along with four reference lines for
     $-U_{t_{lim}}=0.970$, $0.974$, $0.978$, and $0.982$.  
     Notice how $-U_{t_{lim}}=0.982$ has no inner bound for $a=-0.8$, and
     $-U_{t_{lim}}=0.970$ has no solution for $a=1.0$.}  
\end{figure}

The PPI affects slender, intermediate, and wide tori in different ways, so the
choice of torus dimensions is important.  Given a value for $l$, the choice of
surface binding energy, $-U_{t_{lim}}$, sets the overall size of the
disk.  Figure \ref{utlim} illustrates how different choices of $-U_{t_{lim}}$
determine the width of the torus, here shown for $l = 4.5$ and a range of
choices for $a$.  This illustration reinforces the notion that the torus is
created by ``filling" a local minimum in the energy surface; it also
shows that an inner bound for a torus may not exist as $a$ is made increasingly
negative, and the reason for this is best seen in the next figure.

\begin{figure}[ht] 
     \epsscale{1.0}
     \plotone{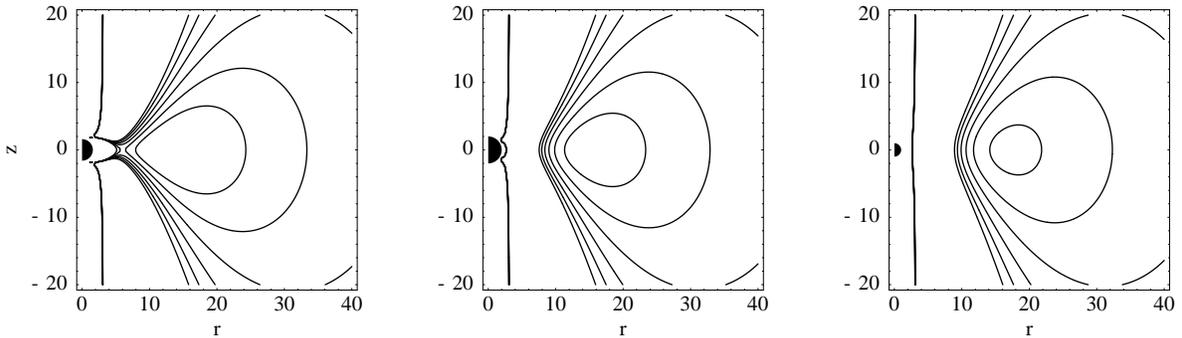}
     \caption{\label{isobars1} Equipotentials for $l=4.5$ disks. The left panel
     is for a retrograde torus, with $a = -0.8$, the middle panel is for
     $a = 0$, the right is for a prograde torus, with $a = 1.0$. These choices
     of $l$ and $a$ are the basis for torus models A3 and B3 described in
     \S4.}
\end{figure}

The role of the Kerr parameter $a$ on the disk configuration is illustrated by
Figure \ref{isobars1}, which shows equipotentials for disks with $l=4.5$, and
the same values of $a$ as Figure \ref{utlim}.  The equipotentials are defined by
the Boyer energy parameter $\Phi=\log(-{U_t})$.  This figure presents polar
``slices'' through the axisymmetric disk.  The set of level contours is the same
in each plot, so the effect of $a$ is readily apparent:  the equipotentials
shrink in the prograde case, and expand in the retrograde case, a result
consistent with what was shown in Figure \ref{utlim} for the equatorial plane.
So, for a given choice of surface binding energy, a retrograde torus would be
larger, and a prograde torus would be more compact, than the Schwarzschild case.

In H91 the value of $l=4.5$ was chosen since it corresponded to a bound torus in
the Schwarzschild metric.  Figures \ref{lmb} and \ref{utlim} hint that this
choice of $l$ may not correspond to bound tori for all values of $a$ down to $a
= -1$.  Figure \ref{isobars1} further reinforces this observation, and also
provides a clearer picture:  many of the contours for the $a = -0.8$ case are
open onto the black hole, and the viable range of parameters for which a bound
torus can be constructed is growing smaller.  For choices of $a$ below $a =
-0.8$ virtually all contours are open onto the hole, so an equilibrium initial 
state cannot be constructed.

\begin{figure}[ht]
     \epsscale{1.0}
     \plotone{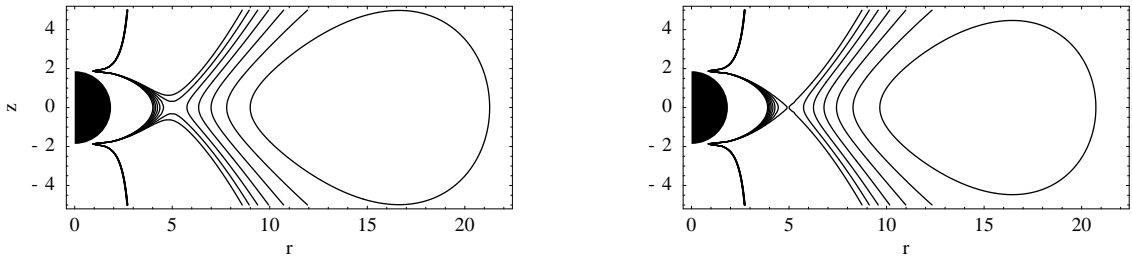}
     \caption{\label{isobars2} Equipotentials for marginal and slightly below 
     marginal prograde tori with $a = -0.5$. 
     The right panel shows the equipotentials for the case where $l = l_{mb}$,
     and the left for  $l = l_{mb}-0.04$. Note how the potential cusp at
     $r = 4.95M$ in the right panel opens up in the left panel.}
\end{figure}

The transition from closed to open contours can best be seen in Figure
\ref{isobars2}, which shows a plot of equipotentials for $l$ at and just below
the marginally bound value for a disk orbiting a Kerr hole with $a=-0.5$ (the
point in Figure \ref{lmb} where the $l_{mb}$ curve passes through $l=4.5$).  The
effect near the cusp is striking:  a small reduction in $l$ below the marginally
bound value opens a pathway to the black hole in the equatorial plane, so that a
disk whose structural parameters give it an inner bound near $r = 4.95\,M$
would not be expected to remain in equilibrium, and any small departure from
equilibrium would trigger accretion. The opening of the potential cusp
can also be accomplished by choosing $a < -0.5$ for an $l = 4.5$ torus,
which explains the shape of the contours in the left panel of Figure
\ref{isobars1}.

Finally the analytic thick disk solution is characterized by a number of derived
parameters.  One is the location of the pressure maximum, $r_{P_{max}}$, which
is obtained by root-finding using the condition $\partial_r \epsilon(r,\pi/2)=0$
since the pressure and energy maxima are coincident.  The angular frequency at
the pressure maximum, $\Omega_{P_{max}} = (g^{t \phi} - l g^{\phi \phi})/(g^{t
t} - l g^{t \phi})$, and the orbital period at the pressure maximum, $T_{orb} =
2\,\pi/\Omega_{P_{max}}$.  Time in the simulations will be reported in terms of
this orbital period.  The boundaries of the disk in the equatorial plane are
$r_{in}$ and $r_{out}$.  Since $r_{in}$ is an input parameter (through the
choice of $U_{t_{lim}}$), a simple plot of $\rho(r,\pi/2,\phi)$ helps locate
$r_{out}$, which is in turn used to specify the required outer boundary of the
radial grid (usually set at about $r_{max} \approx 2\,r_{out}$).

\subsection{Torus Diagnostics}

The Kerr GR hydrodynamics code is used to study the evolution of the
Papaloizou-Pringle instability in thick tori.  In keeping with H91, the
evolution of the instability is characterized by several reduced quantities
extracted from the simulation data.  These include the maximum density
enhancement at mode saturation, $\delta \rho/\rho$, the Fourier power in density
for the azimuthal wavenumbers, $m=1$ and $2$, $q_{sat}$, the angular velocity
distibution parameter at mode saturation, and the mass flux in the equatorial
plane (to help shed light on the relationship between accretion and the growth
of the PPI).  In all models, the calculations are done $20$ times per orbit at
the initial pressure maximum.  The only exception is the mass flux, which is
computed once per orbit.

The density enhancement, $\delta \rho/\rho$,
is obtained by finding the maximum density at in the data arrays, and 
computing $\delta \rho = \rho_{max}{|}_{t=t_{sat}}-\rho_{max}{|}_{t=0}$.

We extract the $m=1$ and $m=2$ Fourier modes by computing 
azimuthal averages using the numerical equivalents of 
\begin{eqnarray}
\Re{k}_m(r)& =&
\int_{0}^{2\pi}{\rho(r,\pi/2,\phi)\,\cos{(m\,\phi)}\,d \phi}\\
\Im{k}_m(r)& =&
\int_{0}^{2\pi}{\rho(r,\pi/2,\phi)\,\sin{(m\,\phi)}\,d \phi}.
\end{eqnarray}
The mode power is then
\begin{equation}
f_m ={1 \over r_{out}-r_{in}}\int_{r_{in}}^{r_{out}}
 {\log_e{\left({(\Re{k}_m(r))}^2 +{(\Im{k}_m(r))}^2\right)}\, d r}.
\end{equation}
A linear fit is performed to the time-sequenced data to extract a mode
growth rate.  For this calculation $r_{in}$ and $r_{out}$ are the
initial values for the inner and outer edges of the torus in the
equatorial plane.  

In H91, the parameter $q_{sat}$ was obtained by a radial power law fit to the
azimuthally averaged angular velocity, $\bar{\Omega}(r) \sim r^{-q}$, at mode
saturation.  This
parameter was used to characterize deviations from a purely Keplerian profile.
Since H91 dealt with Schwarzschild black holes, the angular velocity for the
equilibrium fat disk had a simple form, $\Omega = U^\phi/U^t \sim
g_{tt}\,r^{-2}$, i.e.  the usual Keplerian profile multiplied by the redshift
factor.  It was therefore straightforward to compare the final disk profile
against the Keplerian case.  With Kerr black holes, the equilibrium fat disk has
a more complicated radial dependence, $\Omega = (g^{t \phi} - l g^{\phi
\phi})/(g^{t t} - l g^{t \phi})$.  However, since the aim is to measure a change
in the equatorial angular velocity profile, we adapt the procedure.  As with
H91, we obtain the azimuthally averaged angular velocity using the numerical
equivalent of
\begin{equation}
\bar{\Omega}(r) = {1\over 2\pi} \int_{0}^{2\pi}{V^{\phi}(r,\pi/2,\phi)\,d \phi}.
\end{equation}
A power-law fit $\bar{\Omega}(r) \sim r^{-q}$ is obtained from the slope of a
log-log plot of $\bar{\Omega}(r)$ for $r \in \left(r_{in},r_{out}\right)$ at the
time step corresponding to mode saturation.  We also extract the initial value
$q_{0}$ in an analogous manner, and report the change $\delta\,q =
q_{sat}-q_{0}$ as a measure of the redistribution of angular velocity.

The azimuthally-averaged mass flux in the equatorial plane is  computed
using the numerical equivalent of
\begin{equation}
{\dot{M}}_{eq}(r) = {\sqrt{\gamma(r,\pi/2)}\over 2\pi} 
 \int_{0}^{2\pi}{D(r,\pi/2,\phi)\,V^{\phi}(r,\pi/2,\phi)\,d \phi}.
\end{equation}
In the figures below ${\dot{M}}_{eq}(r)$ is plotted at a radius $r$ lying just
inside the initial inner edge of the torus $r_{in}$ and is used to establish
the presence of a flow of matter towards the black hole.

\section{Results}

Before proceeding with the simulations in the Kerr metric, test runs were
carried out with $a=0$ to establish that this new version of the GR
hydrodynamics solver was able to reproduce the original results in H91.  These
tests were successful and are not elaborated upon here.

\subsection{Parameters for Simulations}

The parameters for the 3D runs in H91, summarized in Table 1, fall into
two broad categories that are distinguished by specific angular
momentum.  The $l=4.5$ models, A3, B3, and C3, are tori with
innermost radii (in the equatorial plane) of $r_{in}=11.0\,M$,
$9.5\,M$, and $8.0\,M$ respectively; these represent slender,
intermediate, and wide tori.  The marginal model, E3,
is a wide torus with $r_{in}=4.4\,M$.  

\begin{table}[ht]
\caption{Disk Parameters in H91.}
\begin{tabular}{lrrrrrl}
 & & & & & &\\
\hline
Model & $l$ & $r_{in}$ & $r_{out}$ & $\Omega$ & $T_{orb}$ & Type\\
\hline
\hline
A3 &  4.50 & 11.0 & 24.7  & 0.0167 & 376 & slender\\
B3 &  4.50 & 9.5  & 37.1  & 0.0167 & 376 & intermediate\\
C3 &  4.50 & 8.0  & 110.0 & 0.0167 & 376 & wide\\
E3 &  3.96 & 4.4  & 79.5  & 0.0313 & 201 & wide (mb)\\
\hline
\end{tabular}
\end{table}

In adapting these models for non-zero Kerr parameters, we are confronted with
the change in torus width with $a$, and also with the restricted range over
which the $l=4.5$ torus remains bound.  To emphasize the role of $a$, we chose
the extreme Kerr limits wherever possible.  Six models were selected, and are
listed in Table 2.  We use the same naming convention as in H91; models are
named according to the choice of the pair of parameters $(l,\,r_{in})$, here
augmented by the Kerr parameter $a$.  Model A3p represents a prograde $l=4.5$
torus in the extreme Kerr limit; the slender A3 torus has expanded to
intermediate size with this change.  Similarly, prograde model B3p has expanded
into a wide torus.  Model B3r represents the ``last" bound retrograde $l=4.5$
torus, for $a=-0.8$; this model is also well below the marginally bound value
for $a=-0.8$ (it could be termed a sub-marginal torus).  The intermediate B3
torus has shrunk considerably with this change in $a$.  The E3 models are more
loosely based on the H91 E3 original; this was done to maintain their nature as
marginal tori, and allowing the values of $r_{in}$ to depart from the H91 value.
Model E3p is a marginal prograde torus with $a=0.5$; the wide marginal E3 torus
has shrunk considerably with this choice of parameters.  Model E3r
is an intermediate, retrograde torus in the extreme Kerr limit.  One additional
model, X3p, was introduced to complete the data set with a prograde
marginal torus that is related to E3p, but whose inner radius lies closer
to the static limit.

\begin{table}[ht]
\caption{\label{Param}Disk Parameters for Numerical Simulations.}
\begin{tabular}{lrrrrrrl}
 & & & & & & &\\
\hline
Model & $a$ & $l$ & $r_{in}$ & $r_{out}$ & $\Omega$ & $T_{orb}$ & Type\\
\hline
\hline
A3p &  1.0 & 4.50 &11.0 & 39.9 & 0.0139 & 452 & intermediate\\
B3p &  1.0 & 4.50 & 9.5 &100.0 & 0.0139 & 452 & wide\\
B3r & -0.8 & 4.50 & 9.5 & 18.3 & 0.0222 & 283 & slender\\
E3p &  0.5 & 3.37 & 4.0 & 17.1 & 0.0550 & 114 & marginal \\
E3r & -1.0 & 4.79 & 8.0 & 57.0 & 0.0164 & 383 & marginal \\
X3p &  0.5 & 3.33 & 3.2 & 23.5 & 0.0594 & 106 & marginal \\
\hline
\end{tabular}
\end{table}

Table 3 is a summary of the simulation results.  The growth rate of the
first and second Fourier modes is given (in units of $\Omega$), as is the
time at which mode saturation occurred (in units of $T_{orb}$), the change in
power-law exponent for the best-fit to the azimuthally-averaged angular
velocity at mode saturation, and the density enhancement at mode
saturation.  Mode saturation is defined as the time at which the first
Fourier mode reaches its maximum amplitude.  For reference the
corresponding results from H91 are given in Table 4.

\begin{table}[ht]
\caption{Three-Dimensional Torus Simulations with $\gamma=5/3$.}
\begin{tabular}{lrrrrrr}
 & & & & & &\\
\hline
Model &  a & $m=1$ & $m=2$ & $t_{sat}$ & $\delta q_{sat}$ & $\delta\,\rho/\rho$\\
\hline
\hline
A3p & 1.0 & 0.096 & 0.063 & 10.0 & -0.15 & 0.57\\
B3p & 1.0 & 0.081 & 0.046 & 13.0 & -0.09 & 0.35\\
B3r &-0.8 & 0.150 & 0.108 & 9.0 & -0.08 & 0.26\\
E3p & 0.5 & 0.027 & 0.015 & 33.0 & -0.01 &-0.01\\
E3r &-1.0 & 0.011 & 0.005 & 20.0 &  0.00 &-0.01\\
X3p & 0.5 & 0.112 & 0.109 &  5.5 &  0.00 &-0.01\\
\hline
\end{tabular}
\end{table}
\begin{table}[ht]
\caption{Results from H91.}
\begin{tabular}{lrrrrl}
 & & & & &\\
\hline
Model & $m=1$ & $m=2$ & $t_{run}$ & $q_{sat}$ & $\delta\,\rho/\rho$\\
\hline
\hline
A3 &  0.175 & 0.112 & 8.5 & 1.80 & 0.94\\
B3 &  0.089 & 0.040 & 16.0 & 1.86 & 0.26\\
C3 &  0.070 & 0.022 & 19.0 & 1.99 & 0.05\\
E3 &  0.022 & 0.021 & 39.0 & 2.00 & 0.006\\
\hline
\end{tabular}
\end{table}

\begin{figure}[ht]
     \epsscale{0.4}
     \plotone{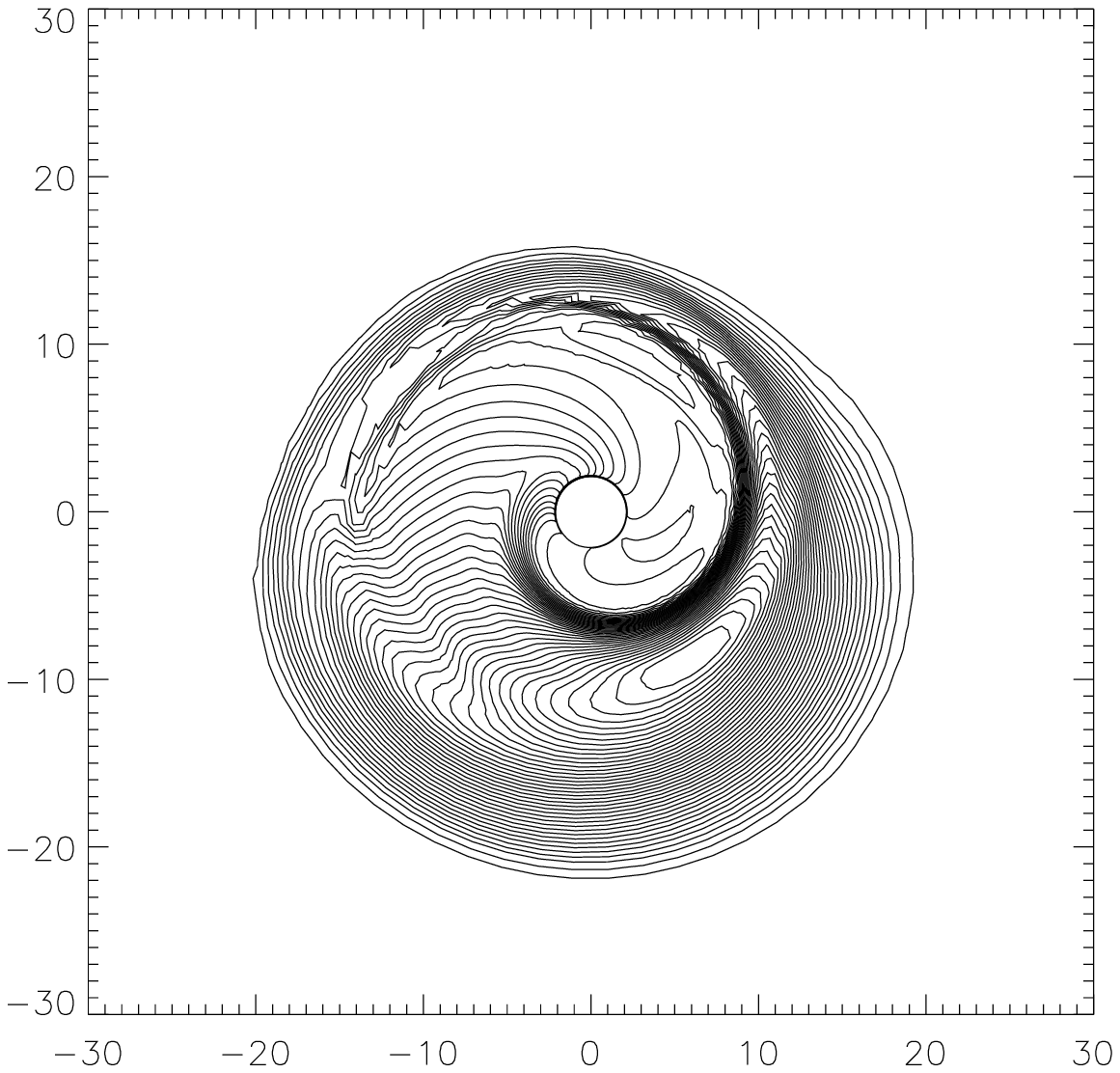}
     \plotone{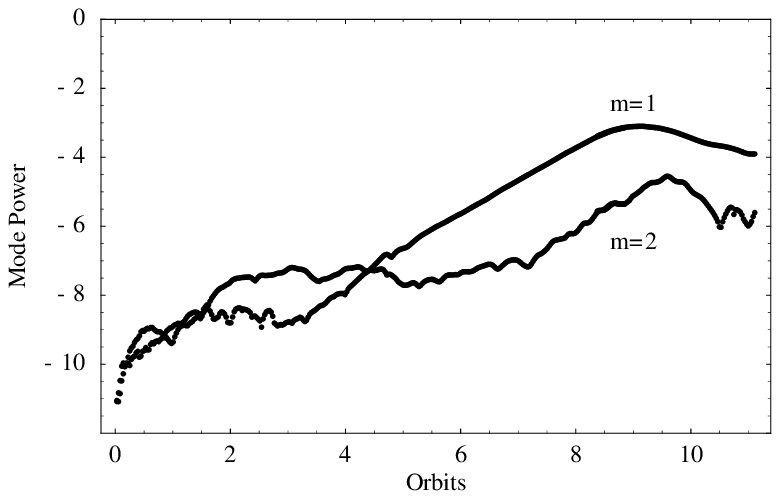}
     \plotone{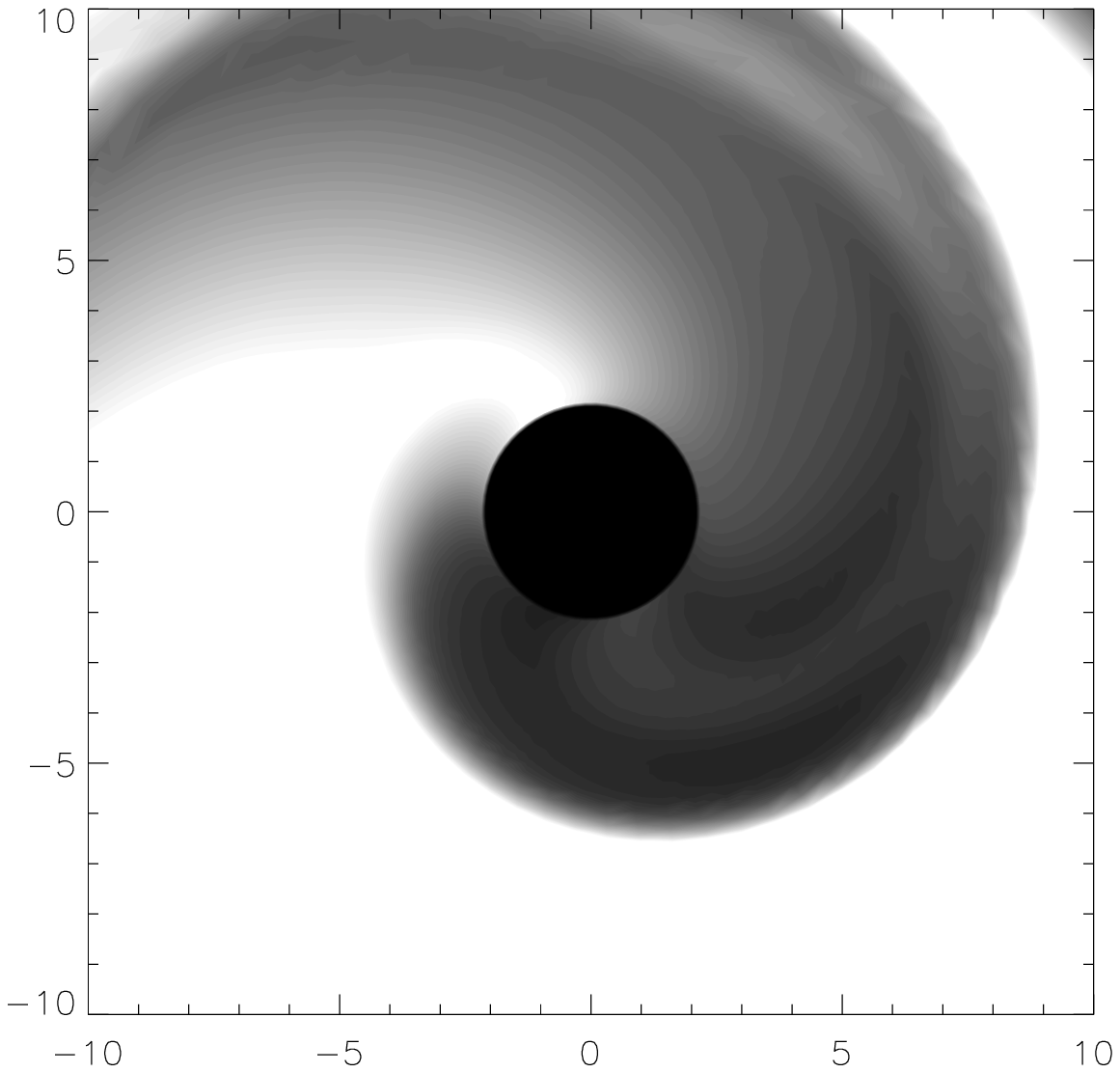}
     \plotone{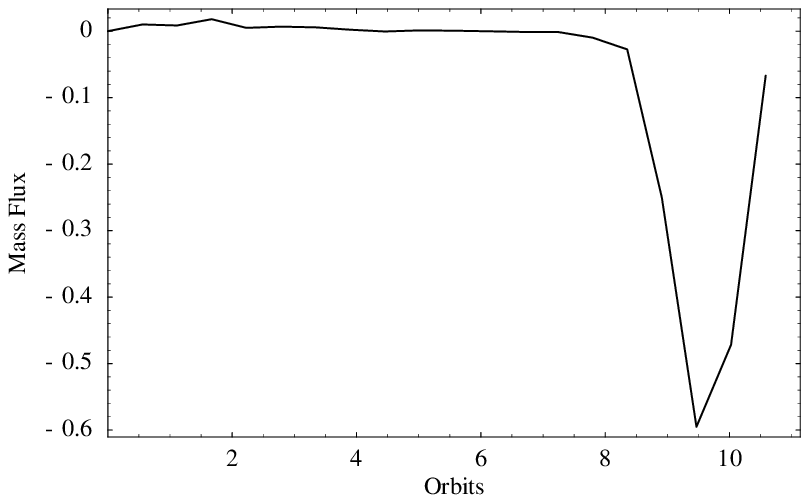}
     \caption{\label{B3r} B3r Model.
     (a) (Top left) Equatorial slice through torus at saturation. Density contours 
     linearly spaced between $\rho_{max}$ and $0.0$.
     (b) (Bottom left) Magnified view of flow near static limit at saturation
         ($\times 4$ density enhancement, linear gray scale). The edge of the 
         central black circle is the static limit.
     (c) (Top right) Mode growth.
     (d) (Bottom right) Mass influx at inner edge of disk. (Black hole rotates in 
     counter-clockwise sense.)}
\end{figure} 
\begin{figure}[ht]
     \epsscale{0.4}
     \plotone{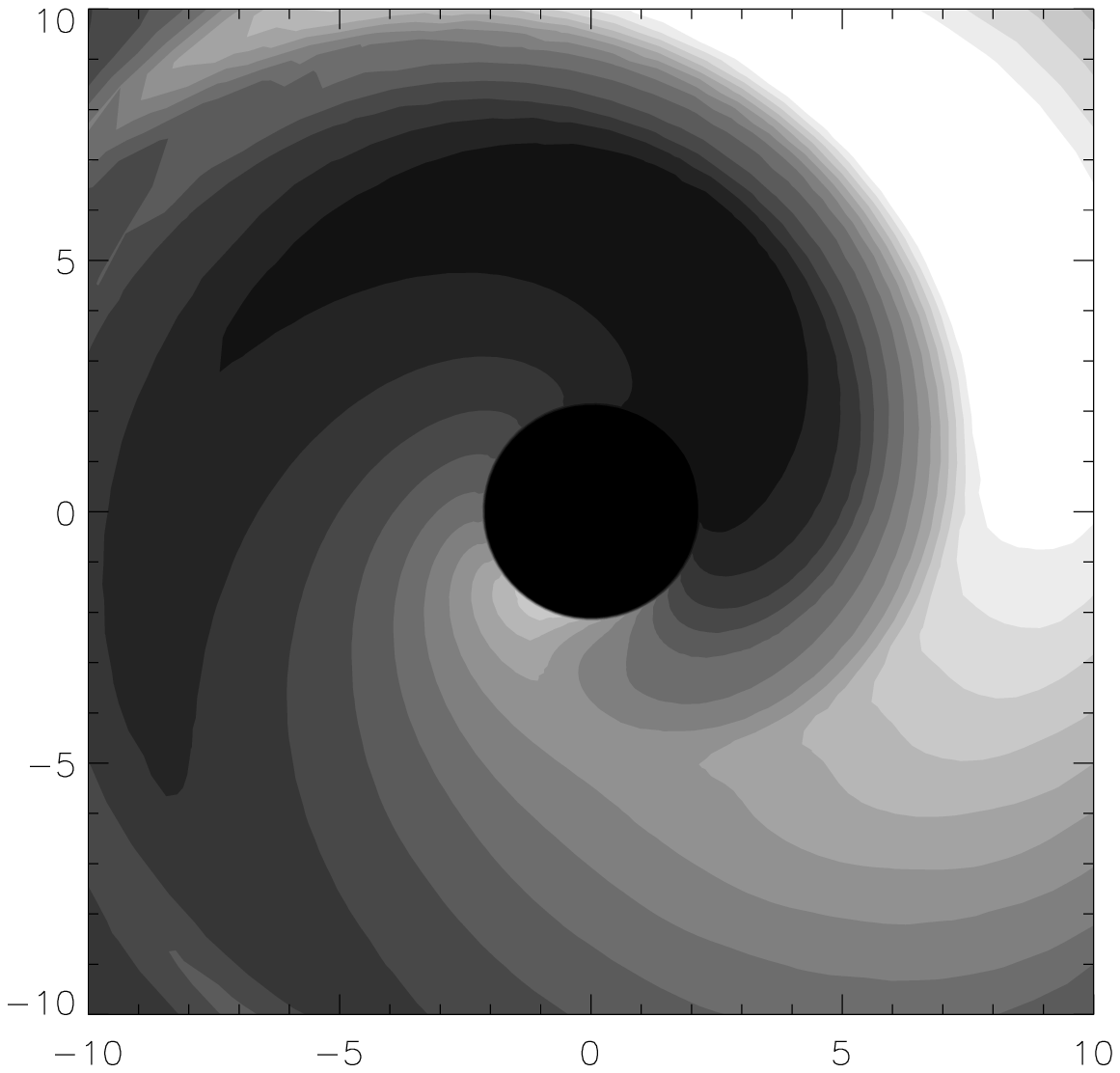}
     \plotone{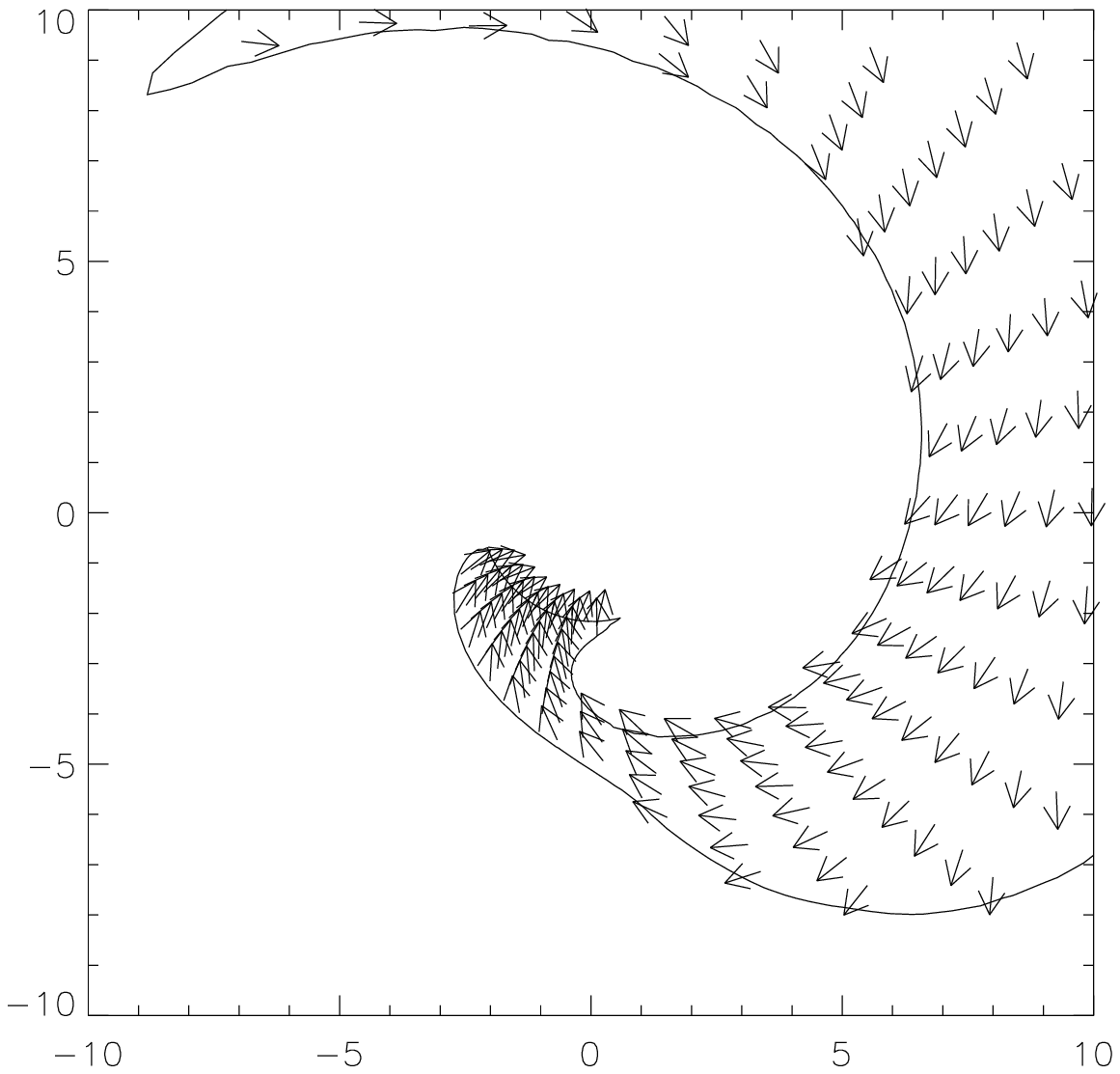}
     \caption{\label{B3r_flow} Direction Vectors for B3r Model. This section through
     the equatorial plane is taken at $10.02\,T_{orb}$ and shows a
     stream of infalling matter that exhibits frame dragging.
     The left panel shows 15 isodensity contours plotted on a linear scale.
     The right panel shows the direction of flow for the matter confined to
     the high density stream. The reversal of flow direction can be clearly
     seen for matter inside $r \sim 4\,M$, showing perhaps the clearest
     evidence for frame dragging.}
\end{figure} 

\subsection{Slender Torus - Model B3r}

Model B3r is a slender torus rotating in the retrograde sense around a Kerr
black hole with $a=-0.8$.  The torus has an initial maximum density of
$\rho_{max_{0}}=0.031$ at $r=12.8\,M$.  Figure \ref{B3r}(a) shows that at mode
saturation a planet has formed, with a density maximum at $r\sim 11\,M$ and a
fractional density enhancement of $0.26$.  The planet has a crescent shape with
some evidence of a tightly-wrapped outward spiral wave in the lower left
quadrant of panel (a).  A gray-scale enhanced view of the inner region near the
static limit clearly shows that a strong inflow has developed that contacts the
static limit in the upper left quadrant of panel (b); there is also a hint of
frame dragging in the reversal of the sense of the flow near the static limit
(in this figure, as in all subsequent ones, the black hole rotates in the
counter-clockwise sense, and the retrograde torus here rotates in the opposite
sense).  There is also a evidence of a weaker inbound spiral of matter in the
right half of panel (b).  The growth of the PPI modes has two distinct stages,
as seen in Figure \ref{B3r}(c).  The $m=2$ mode grows quickly at the beginning,
then levels off at $2\,T_{orb}$ and remains roughly constant until $7\,T_{orb}$
before increasing again, paralleling the growth of the $m=1$ mode.  The $m=1$
mode growth does not become apparent until approximately $3\,T_{orb}$, at which
point it exhibits strong linear growth to saturation at $9\,T_{orb}$.  Up to
$4\,T_{orb}$, the $m=2$ mode is dominant, but lags the $m=1$ mode in the later
stages of the simulation.  The development of the accretion flow is clarified in
Figure \ref{B3r}(d), which shows the mass influx in the equatorial plane inside
the inner edge of the torus.  There is no significant accretion until
$8\,T_{orb}$.  The accretion rate reaches a maximum shortly after mode
saturation.  This feature, as will be seen in subsequent runs, is associated
with all solutions that yield planets.


The effect of frame dragging is illustrated in Figure \ref{B3r_flow},
which is an equatorial slice at $10.02\,T_{orb}$, one orbit after
saturation.   The left panel of the figure is a density plot with 15
contours, chosen to highlight the stream of high-density material
flowing from the inner edge of the grid into the hole.  The right panel
shows momentum direction vectors inside the flow stream outlined by the
overlaid contour.  The expected reversal in azimuthal flow direction
due to frame dragging begins at $r \sim 4\,M$.  The direction becomes
progressively more prograde as the flow approaches the the inner edge
of the grid, which lies at $2.05\,M$ (just outside the static limit).
This reversal, from retrograde to prograde, produces the dog-leg
pattern seen in the density plot as the flow approaches the static
limit.  A similar pattern is visible for model E3r (see below).


\begin{figure}[ht]
     \epsscale{0.4}
     \plotone{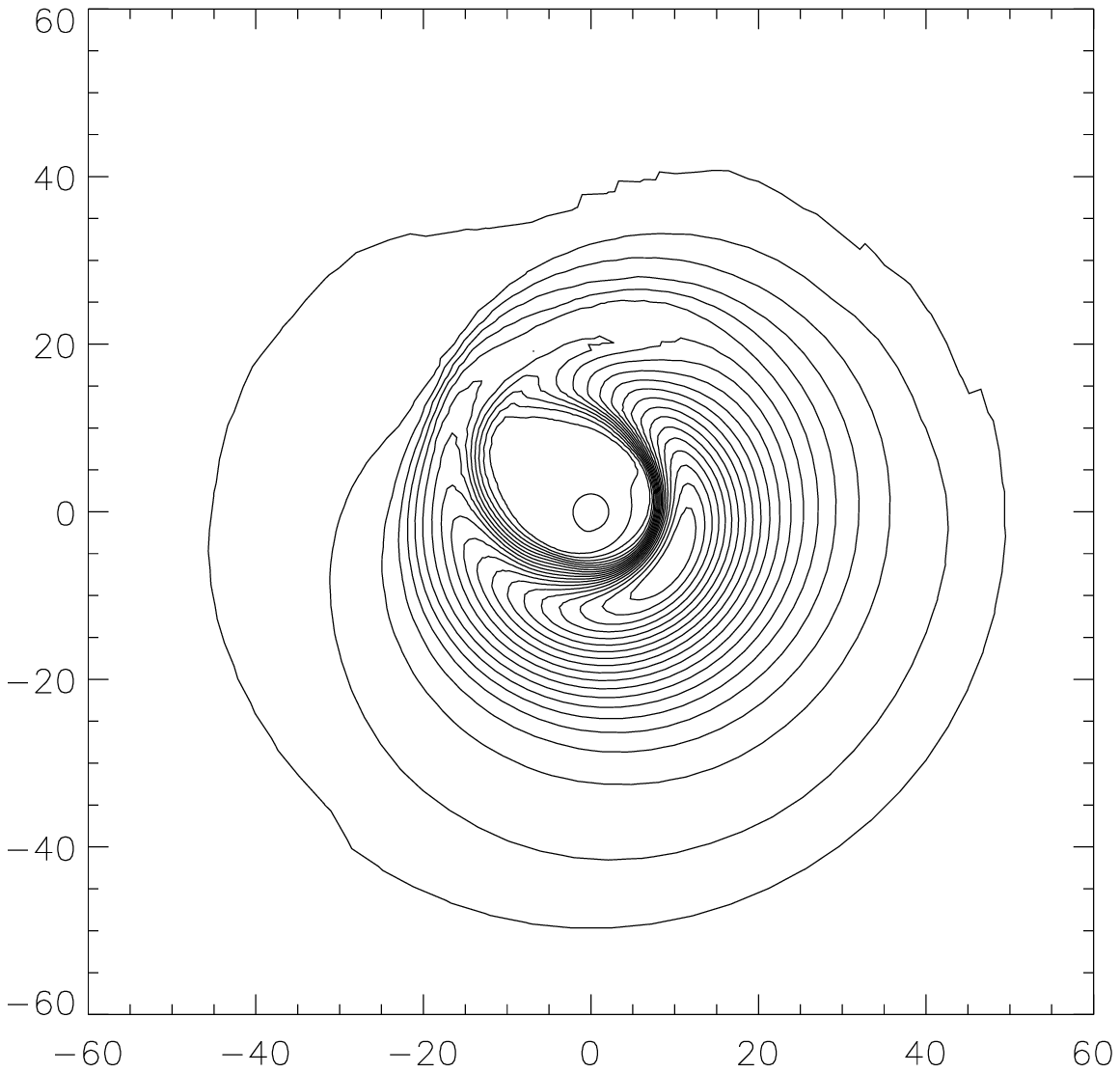}
     \plotone{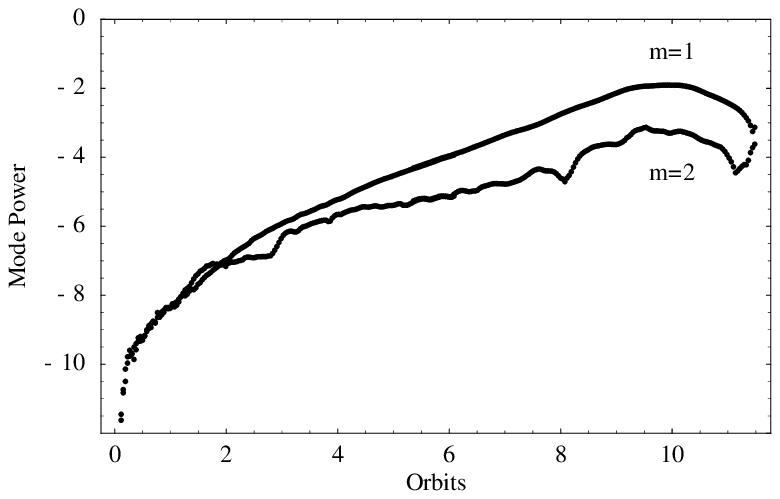}
     \plotone{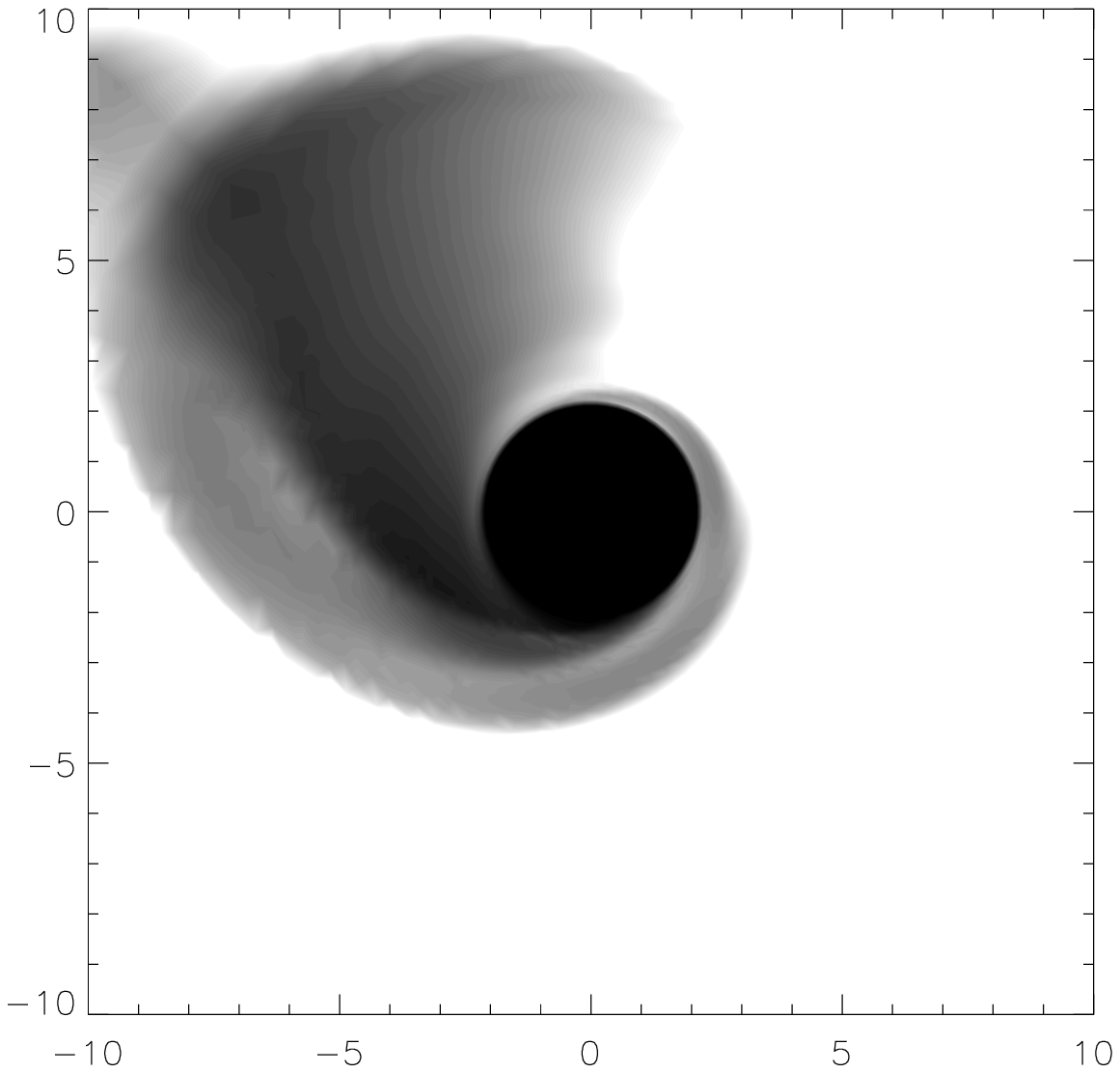}
     \plotone{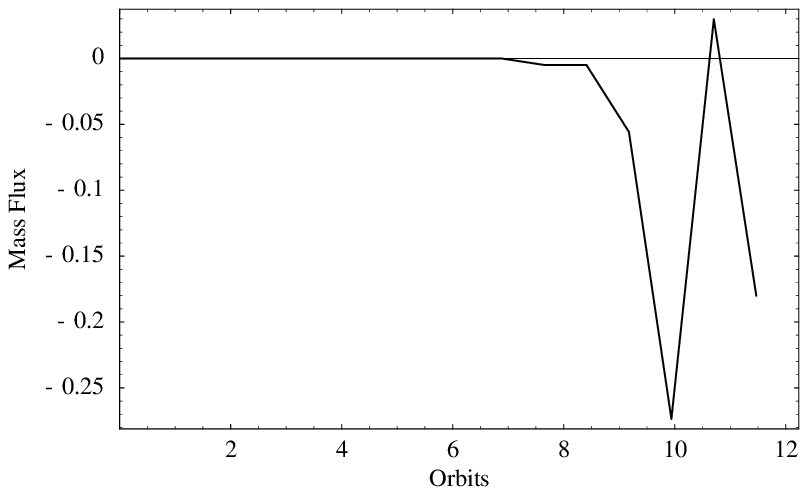}
     \caption{\label{A3p} A3p Model.  
     (a) (Top left) Equatorial slice through torus at saturation. Density contours 
     linearly spaced between $\rho_{max}$ and $0.0$.
     (b) (Bottom left) Magnified view of flow near static limit at saturation
         ($\times 50$ density enhancement, linear gray scale). The edge of the 
         central black circle is the static limit.
     (c) (Top right) Mode growth.
     (d) (Bottom right) Mass influx at inner edge of disk. (Black hole rotates in 
     counter-clockwise sense.)}
\end{figure} 

\subsection{Intermediate Torus - Model A3p}

Model A3p is an intermediate torus rotating in the prograde sense around a Kerr
black hole with $a=1.0$.  The torus has an initial maximum density of
$\rho_{max_{0}}=0.23$ at $r=17.2\,M$.  Figure \ref{A3p}(a) shows that at mode
saturation a planet has formed, with a density maximum at $r\sim 11\,M$ and a
fractional density enhancement of $0.57$.  The planet has a narrow crescent
shape with an outward spiral wave visible in the upper left quadrant of panel
(a).  A gray-scale highly enhanced ($\times 50$) view of the inner region near
the static limit clearly shows that a weak inspiral of material is underway at
saturation; however, this is by far the weakest of the inflows in the set of
planet-forming simulations.  The growth of the PPI modes is strong from the
outset, as seen in Figure \ref{A3p}(c).  The $m=1$ and $m=2$ modes track one
another, with the $m=1$ mode growing at a slightly greater rate.  The $m=1$ mode
is also less erratic than the $m=2$ mode.  Both modes exhibit strong linear
growth up to saturation at $10\,T_{orb}$.  A small amount of inflow is present
at saturation, as can be seen in Figure \ref{A3p}(d), which shows the mass
influx in the equatorial plane inside the inner edge of the torus.  The inflow
of matter is quiescent until $8\,T_{orb}$, and reaches a maximum at mode
saturation.  The apparent reversal of the flow after $10\,T_{orb}$, combined
with the lack of evidence for substantial inflow in panel (b), suggests that
the planet is settling down to a new equilibrium state, and that this apparent
outward flow of matter is simply indicative of the reorganization of matter
within the disk.

\subsection{Wide Torus - Model B3p}

\begin{figure}[ht]
     \epsscale{0.4}
     \plotone{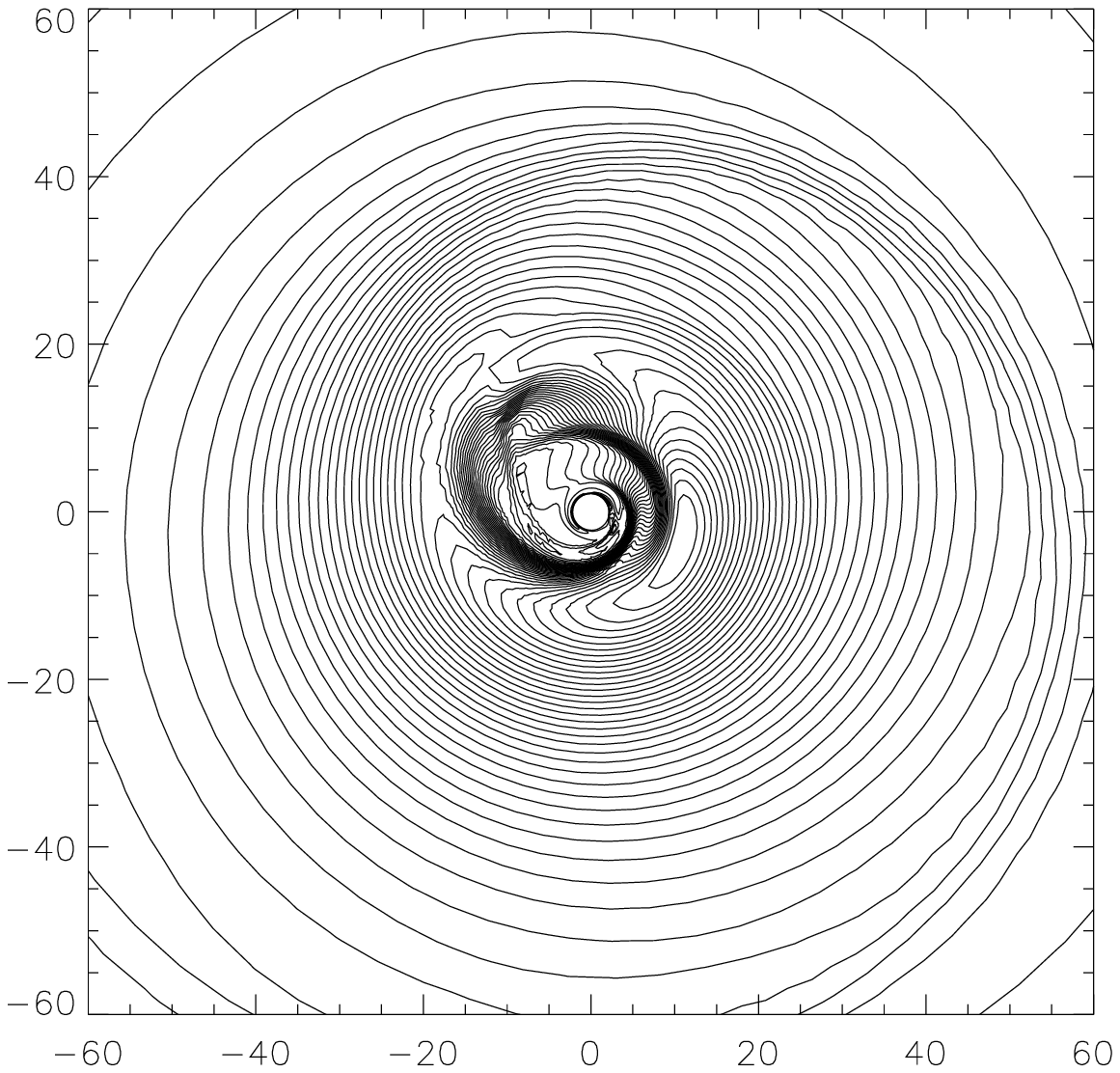}
     \plotone{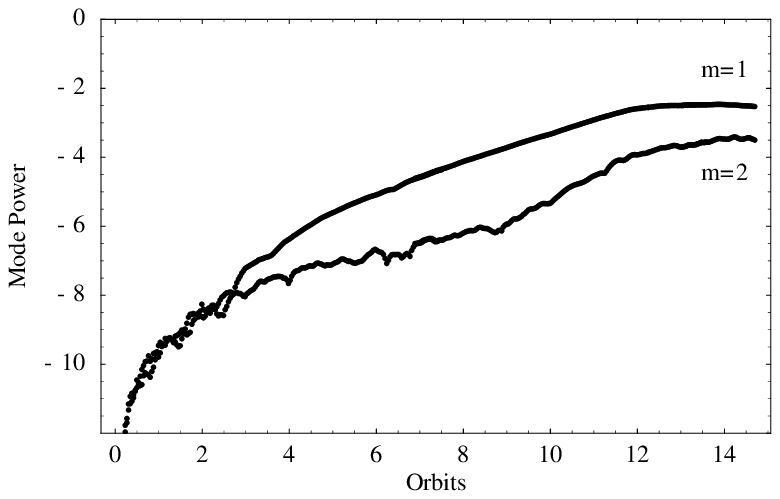}
     \plotone{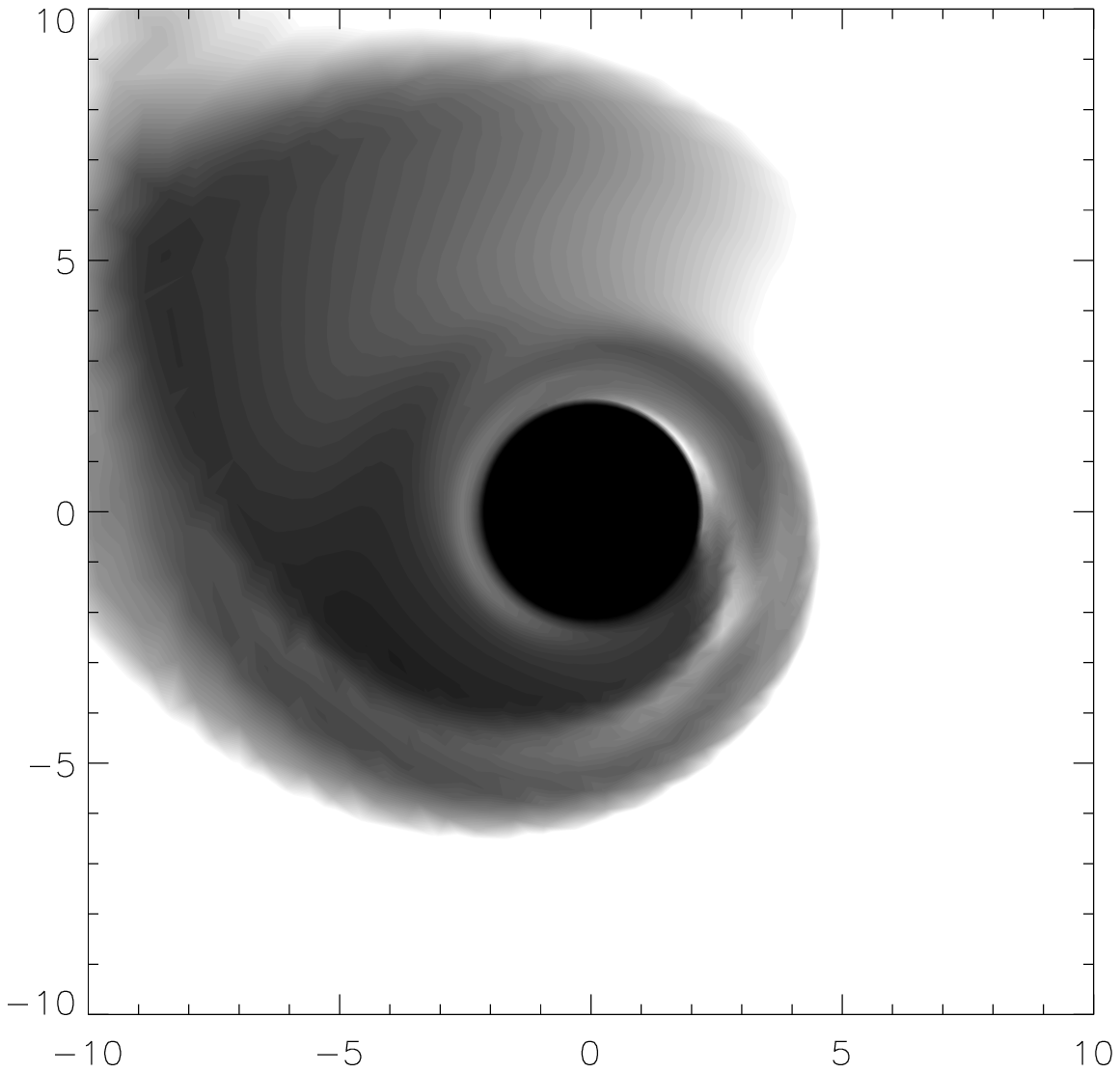}
     \plotone{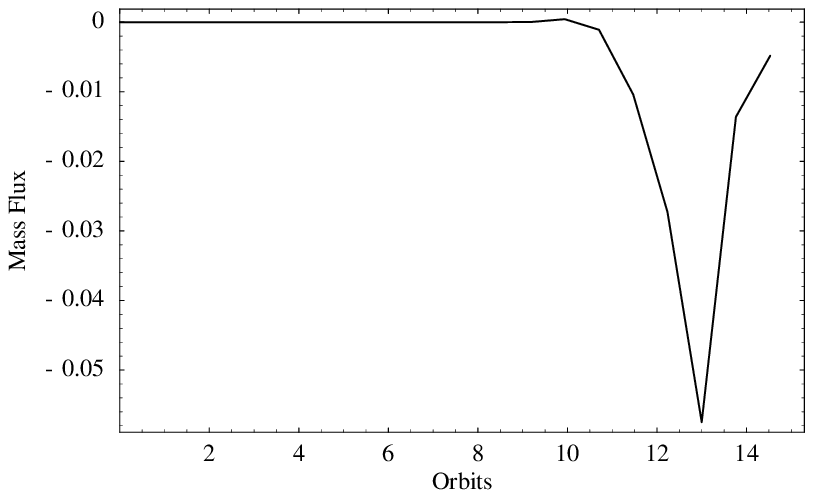}
     \caption{\label{B3p}B3p Model.
     (a) (Top left) Equatorial slice through torus at saturation. Density contours 
     linearly spaced between $\rho_{max}$ and $0.0$.
     (b) (Bottom left) Magnified view of flow near static limit at saturation ($\times 4$ 
     density enhancement). The edge of the central black circle is the static 
     limit.
     (c) (Top right) Mode growth.
     (d) (Bottom right) Mass influx at inner edge of disk. (Black hole rotates 
     in counter-clockwise sense.)}
\end{figure}

Model B3p is a wide torus rotating in the prograde sense around a Kerr black
hole with $a=1.0$.  The torus has an initial maximum density of
$\rho_{max_{0}}=0.72$ at $r=17.2\,M$.  Although the disk extends to $r \approx
90 M$, most of the matter is concentrated inside $r \approx 30 M$.  Figure
\ref{B3p}(a) shows that at mode saturation a planet has formed, with a density
maximum at $r\sim 11\,M$ and a fractional density enhancement of $0.35$.  The
planet has a tightly-wrapped crescent shape with a noticeable outward spiral
wave in the upper right quadrant of panel (a).  A gray-scale enhanced view of
the inner region near the static limit clearly shows that an inflow has
developed that contacts the static limit in the lower right quadrant of panel
(b).  There is also a hint of frame dragging of the flow near the static limit
where dense material appears to smear out into a ring.  A weaker inbound spiral
of matter in the upper half of panel (b) is also feeding the inner ring of
matter.  The growth of the PPI modes is strong from the outset, as seen in
Figure \ref{B3p}(c).  The $m=1$ and $m=2$ modes track one another, with the
$m=1$ mode growing a a slightly greater rate.  The $m=1$ mode is also less
erratic than the $m=2$ mode.  Both modes exhibit strong linear growth up to
saturation at $\sim 13\,T_{orb}$; the saturation peak is very broad, making a
precise determination of $t_{sat}$ difficult.  As with the previous two models,
inflow is triggered at saturation, as can be seen in Figure \ref{B3p}(d), which
shows the mass influx in the equatorial plane inside the inner edge of the
torus.  The flow of matter is quiescent until $11\,T_{orb}$, and reaches a
maximum at $13\,T_{orb}$, again corresponding with mode saturation.

\subsection{Marginal Torus - Model E3p}

\begin{figure}[ht]
     \epsscale{0.4}
     \plotone{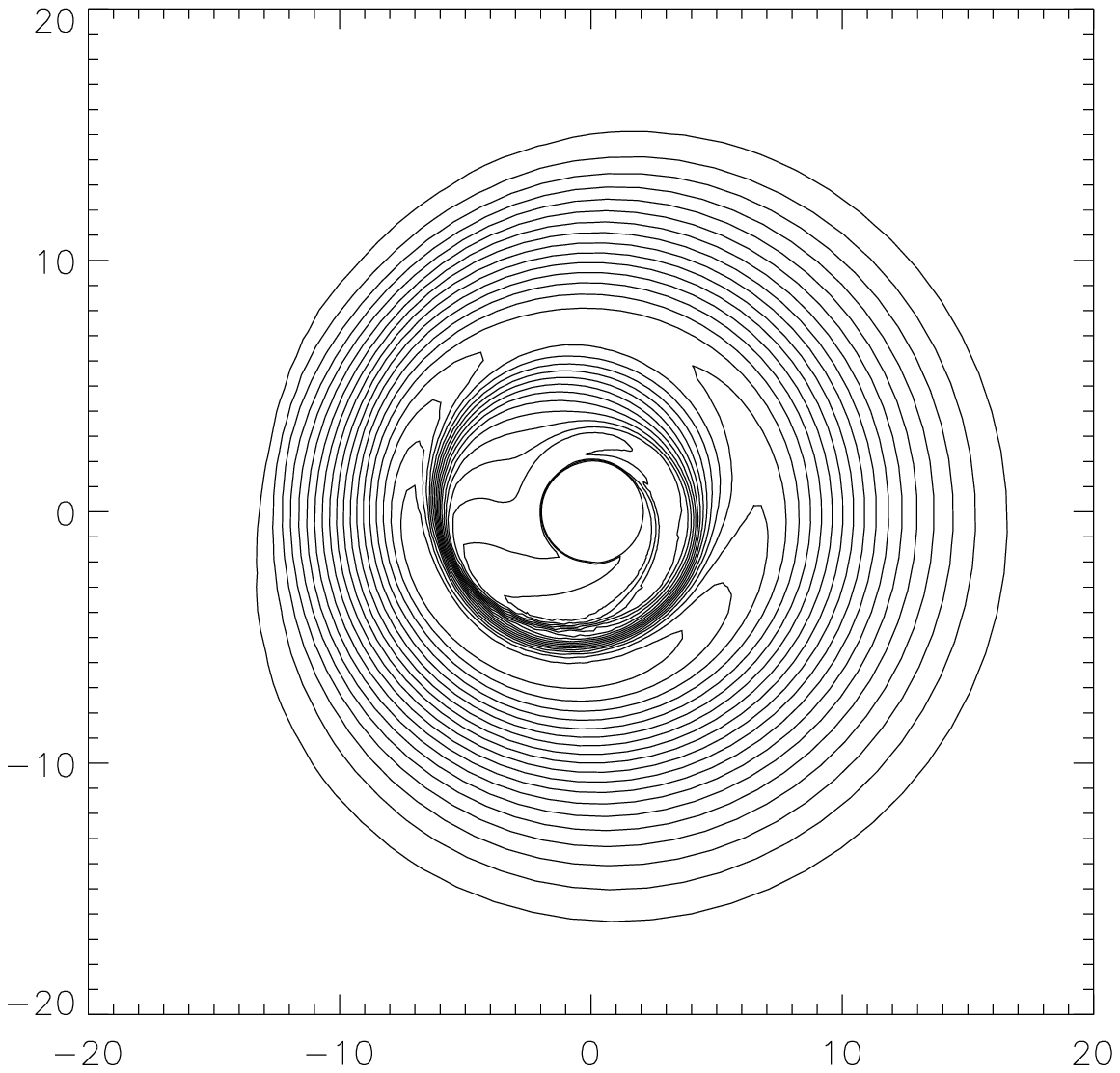}
     \plotone{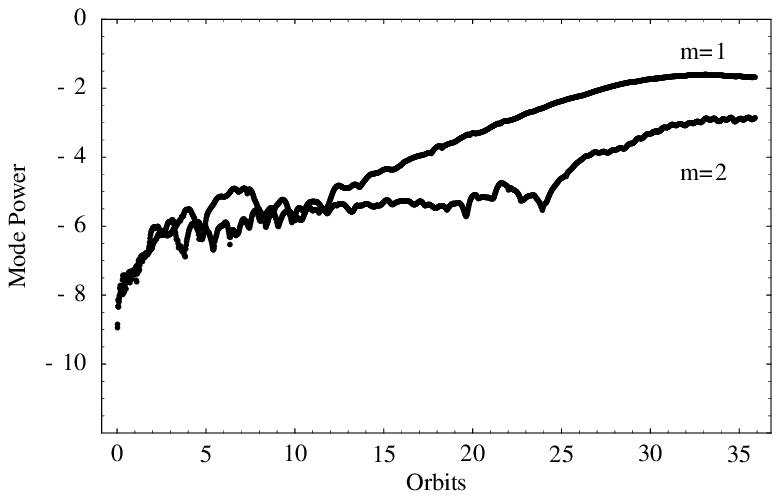}
     \plotone{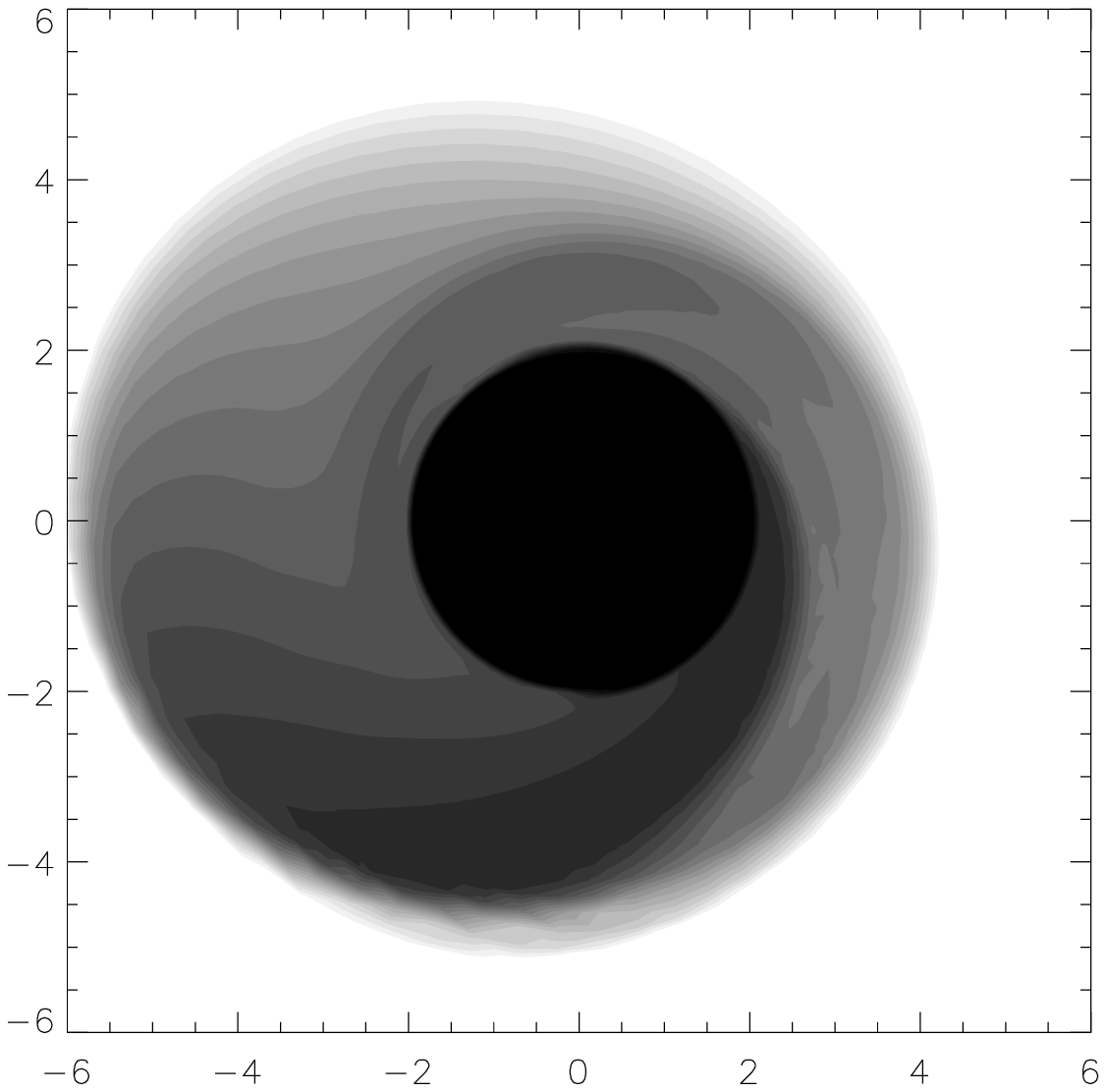}
     \plotone{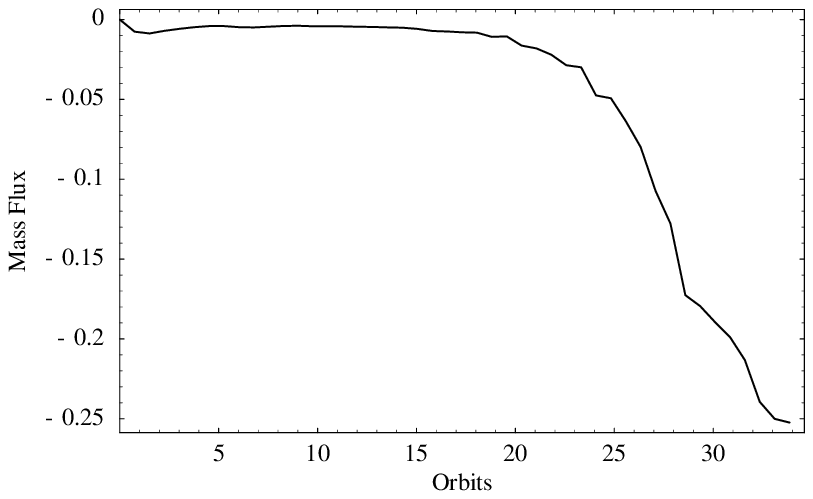}
     \caption{\label{E3p}E3p Model. 
     (a) (Top left) Equatorial slice through torus at saturation. Density contours 
     linearly spaced between $\rho_{max}$ and $0.0$.
     (b) (Bottom left) Magnified view of flow near static limit at saturation ($\times 2$  
     density enhancement). The edge of the central black circle is the static 
     limit.
     (c) (Top right) Mode growth.
     (d) (Bottom right) Mass influx at inner edge of disk. (Black hole rotates 
     in counter-clockwise sense.)}
\end{figure}

Model E3p is a marginal torus rotating in the prograde sense around a Kerr black
hole with $a=0.5$.  The torus has an initial maximum density of
$\rho_{max_{0}}=0.86$ at $r=6.8\,M$.  Figure \ref{E3p}(a) shows that at mode
saturation an elongated planet has formed, with a density maximum at $r\sim
6\,M$ and a fractional density decrease of $0.01$.  A gray-scale view ($\times
2$ enhancement) of the inner region near the static limit clearly shows that an
inflow has developed that contacts the static limit in the upper right quadrant
of panel (b).  Figure \ref{E3p}(c) shows that the $m=1$ mode undergoes steady
growth from the outset, with some variability in the first $10$ orbits.  The
$m=2$ mode tracks the $m=1$ mode until $12\, T_{orb}$, where it flattens out
until $25\, T_{orb}$, then grows again until the $m=1$ mode saturates at $33\,
T_{orb}$.  Perhaps the most striking feature of this model can be seen in the
mass influx curve, Figure \ref{E3p}(d), again plotted for a point in the
equatorial plane inside the inner edge of the disk.  The transition between
quiescence and accretion is less sharply defined than in other models, and a
strong inflow begins after $\sim 25\,T_{orb}$, roughly corresponding to the
point where the $m=2$ mode begins to grow again.  Compared to the other models,
the amount of accretion in this model is remarkable; panels (a) and (b) clearly
show very prominent in-bound spirals of matter.

\subsection{Marginal Torus - Model E3r}

\begin{figure}[ht]
     \epsscale{0.4}
     \plotone{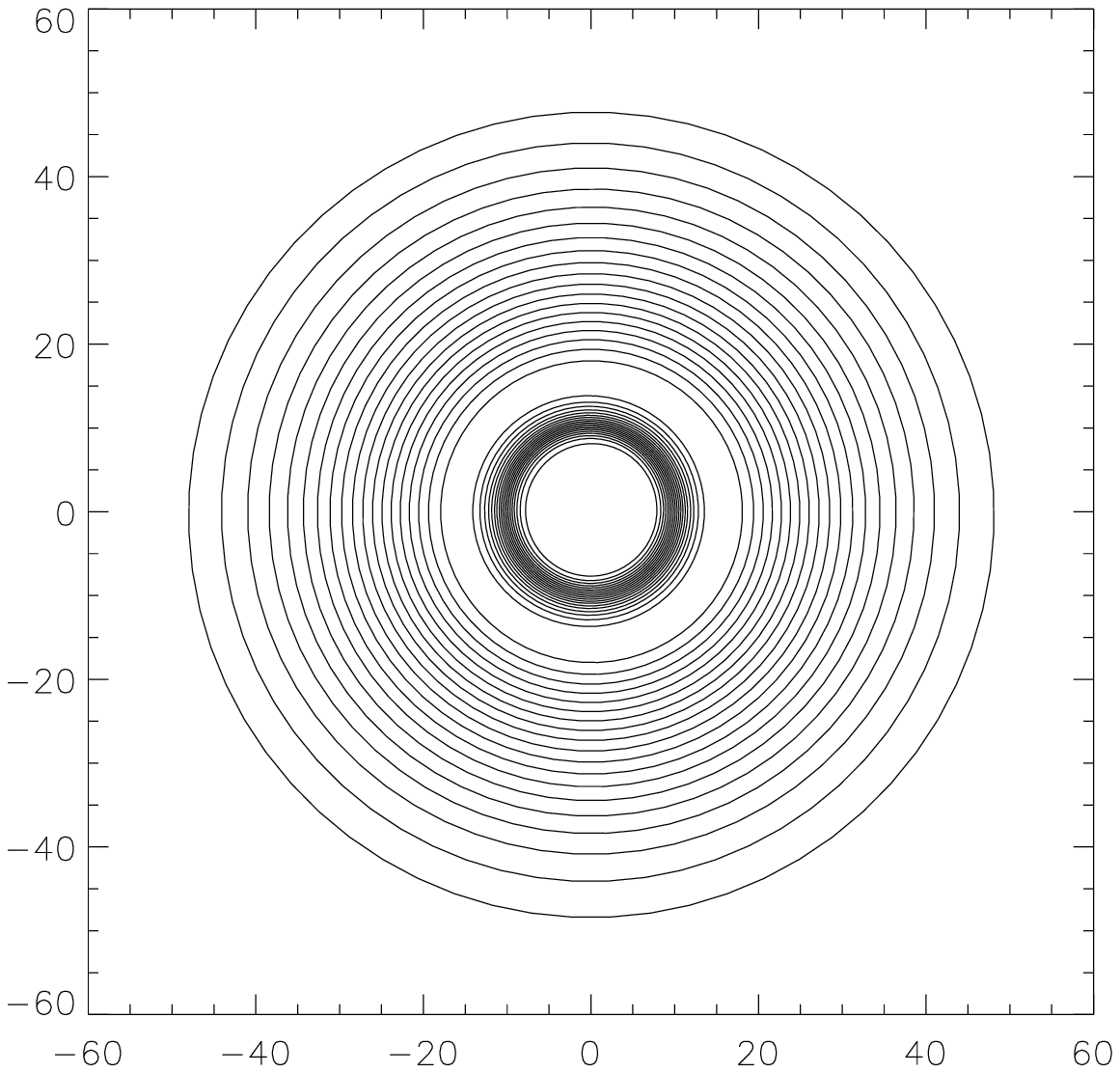}
     \plotone{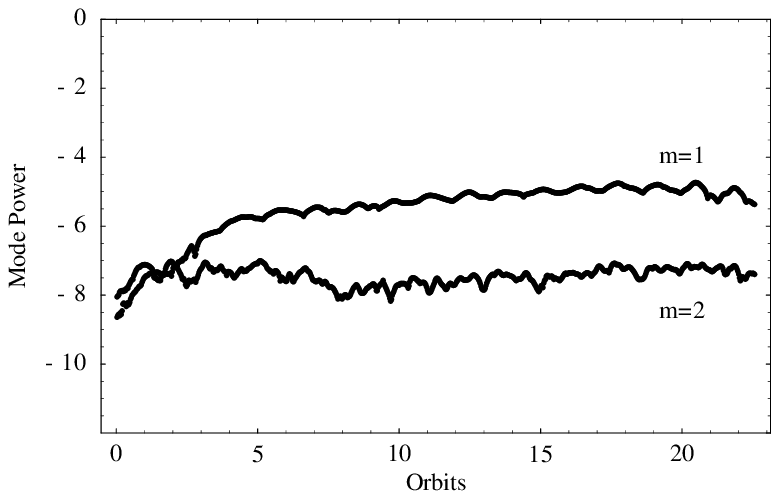}
     \plotone{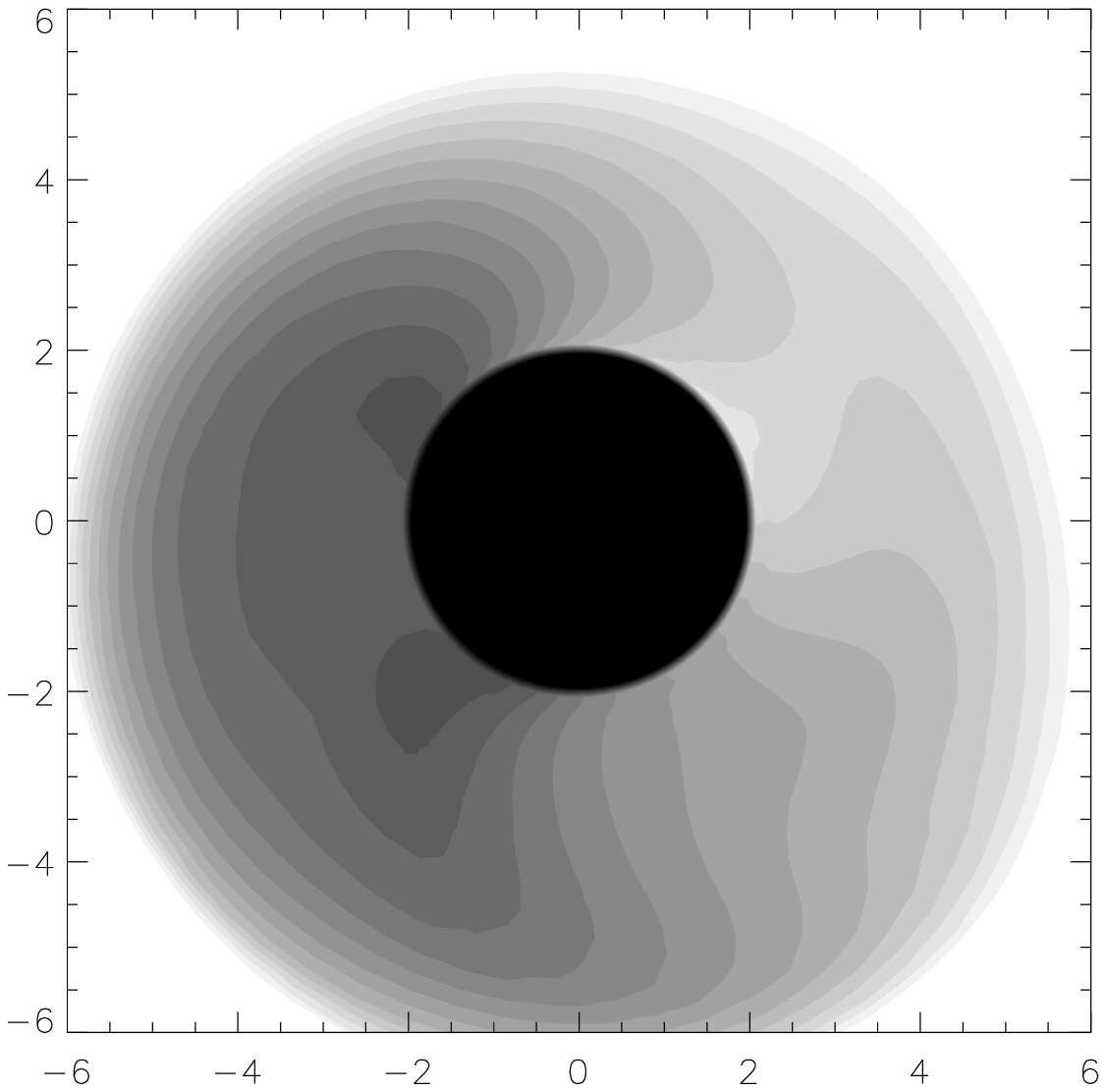}
     \plotone{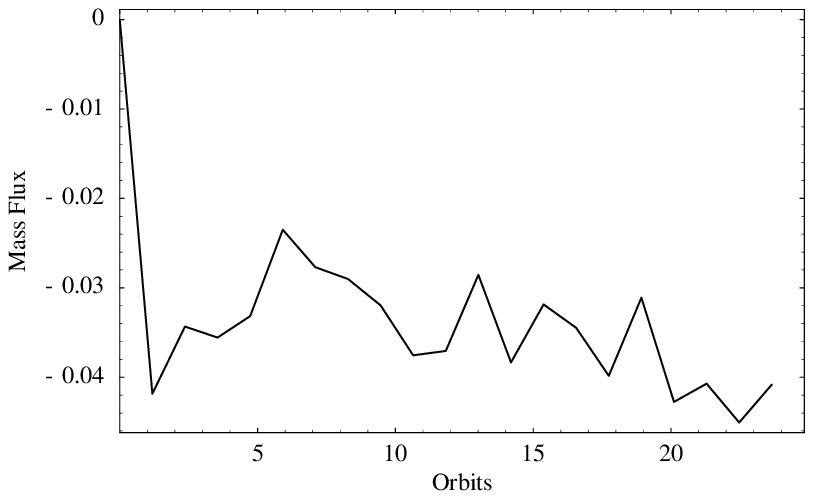}
     \caption{\label{E3r}E3r Model.
     (a) (Top left) Equatorial slice through torus at saturation. Density contours 
     linearly spaced between $\rho_{max}$ and $0.0$.
     (b) (Bottom left) Magnified view of flow near static limit at saturation ($\times 100$ 
     density enhancement). The edge of the central black circle is the static 
     limit.
     (c) (Top right) Mode growth.
     (d) (Bottom right) Mass influx at inner edge of disk. (Black hole rotates in 
     counter-clockwise sense.)}
\end{figure}

Model E3r is a marginal torus rotating in the retrograde sense around a
Kerr black hole with $a=-1.0$.  The torus has an initial maximum density of
$\rho_{max_{0}}=0.43$ at $r=15.7\,M$.  Figure \ref{E3r}(a) shows that at mode
saturation the disk remains largely unperturbed, with an insignificant
fractional density decrease of $0.01$.  Figure \ref{E3r}(b) shows a gray-scale
enhanced view of the inner region near the static limit clearly shows that an
inflow has developed that contacts the static limit in the upper right quadrant
of panel (b); it must be noted that this inflow is extremely weak ($\times 100$
enhancement) yet it bears some similarity to the other retrograde model, B3r, in
that the stream of matter exhibits a change in flow direction near the static
limit.  The growth of the PPI modes is weak, with the $m=1$ mode levelling off
quickly after a brief initial growth, and the $m=2$ mode showing little
activity, as seen in Figure \ref{E3r}(c).  The $m=1$ mode reaches a very broad,
weak maximum at $\sim 20\,T_{orb}$, making a precise determination of $t_{sat}$
difficult.  Unlike the previous models, a noticeable amount of inflow is present
from the outset as can be seen in Figure \ref{E3r}(d), which shows the mass
influx in the equatorial plane inside the inner edge of the torus.  Since this
model is marginal, the early onset of an inflow is to be expected.  In
comparing panels (c) and (d), we see that the inflow is established before any
significant mode growth has occured; inflow levels off to a more or less
constant value in the first 5 orbits; the mode growth also levels off in this
period.  This would suggest that early accretion is an effective mechanism for
inhibiting mode growth.  This observation is strengthened by the last model to
be considered, X3p.

\subsection{Marginal Torus - Model X3p}

\begin{figure}[ht]
     \epsscale{0.4}
     \plotone{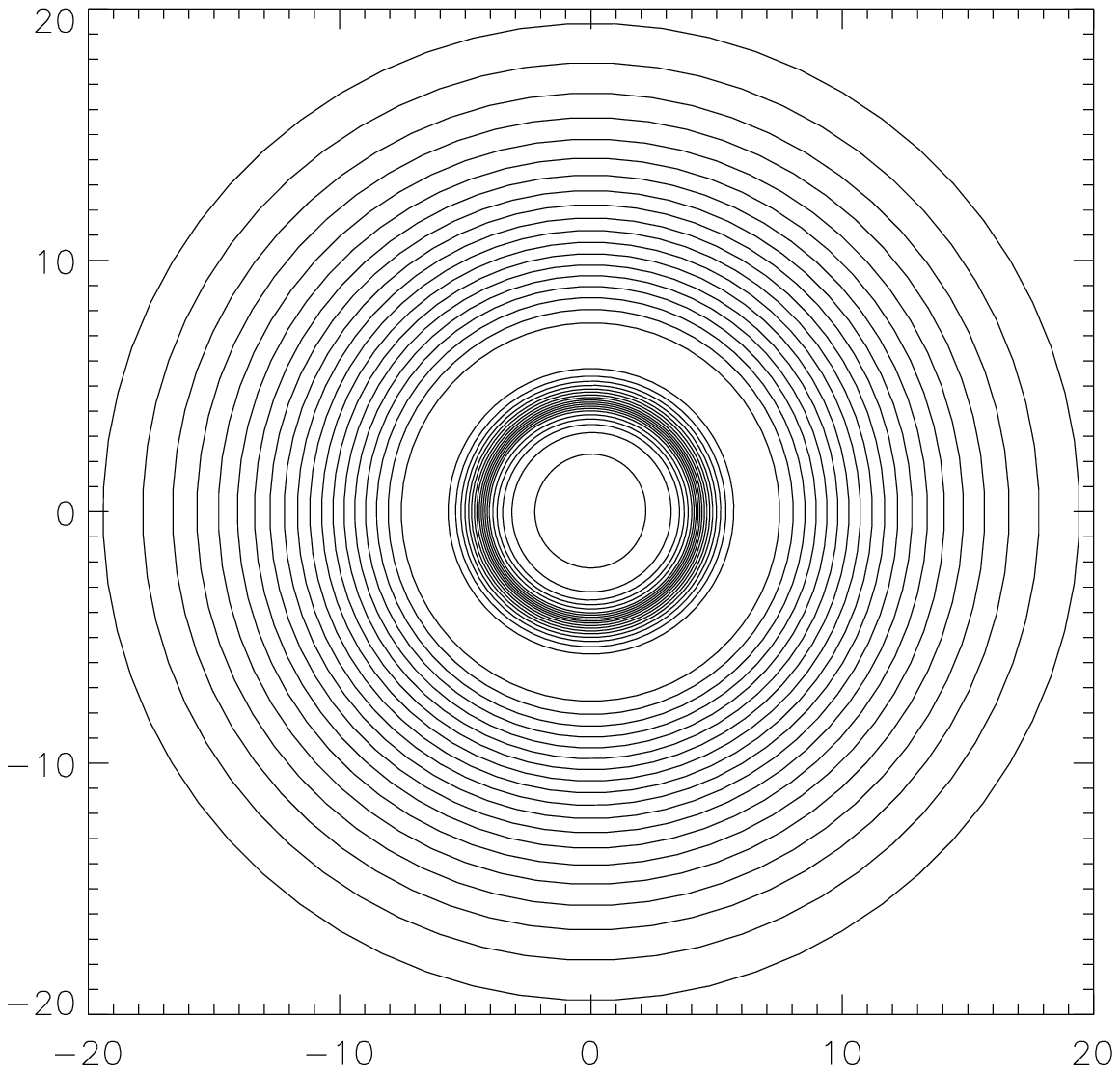}
     \plotone{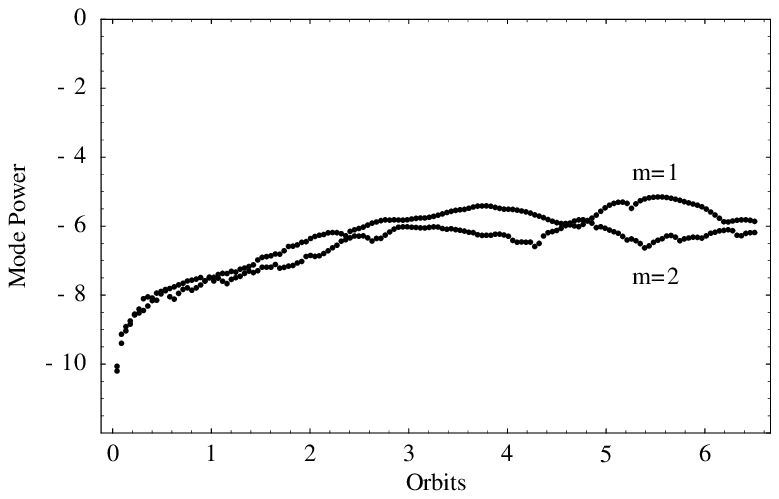}
     \plotone{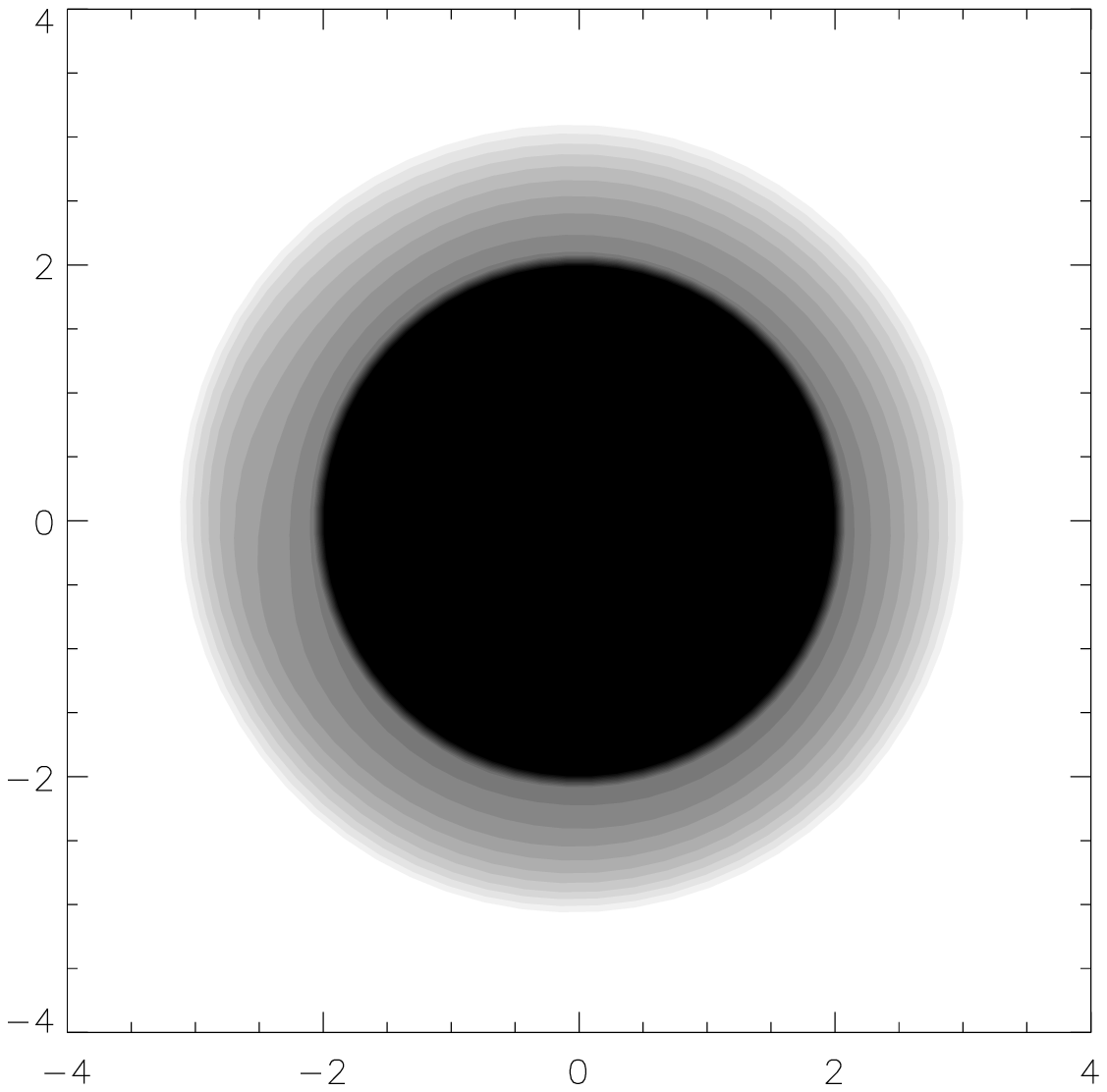}
     \plotone{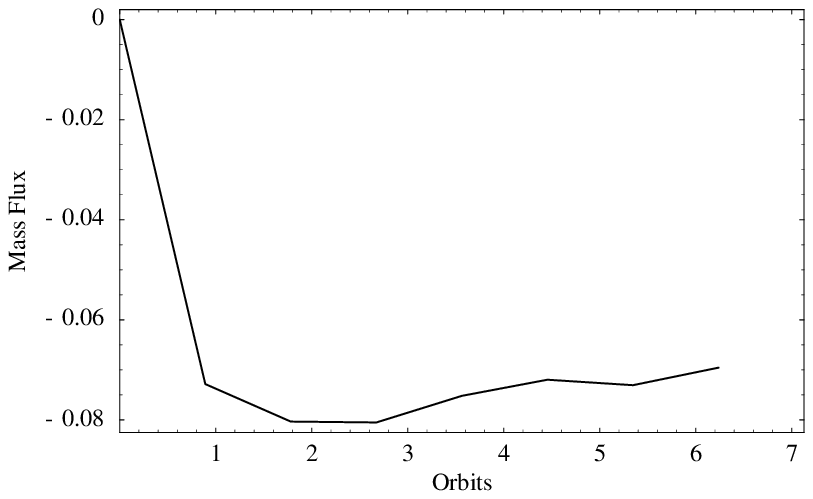}
     \caption{\label{X3p}X3p Model.
     (a) (Top left) Equatorial slice through torus at saturation. Density contours
     linearly spaced between $\rho_{max}$ and $0.0$.
     (b) (Bottom left) Magnified view of flow near static limit at saturation ($\times 10$ 
     density enhancement). The edge of the central black circle is the static 
     limit.
     (c) (Top right) Mode growth.
     (d) (Bottom right) Mass influx at inner edge of disk. (Black hole rotates in 
     counter-clockwise sense.)}
\end{figure}

Model X3p is a marginal torus rotating in the prograde sense
around a Kerr black hole with $a=0.5$.  The torus has an initial maximum density
of $\rho_{max_{0}}=1.58$ at $r=6.4\,M$.  Figure \ref{X3p}(a) shows that at mode
saturation the disk remains largely unperturbed, with an insignificant
fractional density decrease of $0.01$.  Figure \ref{X3p}(b) shows a gray-scale
enhanced view of the inner region near the static limit clearly shows that a
highly symmetric, weak inflow has developed.  The growth of the PPI modes is
weak, with both modes tracking one another, as seen in Figure \ref{X3p}(c).
Although Table 3 would suggest that these modes have a vigorous growth rate,
their maximum amplitude is comparable to that of model E3r, and the outcome is
very similar to that model, which is interesting since model X3p is much more
closely related, in terms of its structural parameters, to model E3p.  The $m=1$
mode reaches a very broad, weak maximum at $\sim 5.5\,T_{orb}$.  As with model
E3r, a noticeable amount of inflow is present from the outset as can be seen in
Figure \ref{X3p}(d), which shows the mass influx in the equatorial plane inside
the inner edge of the torus.  As with E3r, this model is marginal, and the
early onset of an inflow is to be expected.  In comparing panels (c) and (d), we
see that the inflow is established before any significant mode growth has
occured; inflow levels off to a more or less constant value in the first 4
orbits; the mode growth also levels off in this period.  As with model E3r,
early accretion appears to be an effective mechanism for inhibiting mode growth.

\section{Conclusions}

The numerical simulations presented in this paper were undertaken to
test the hydrodynamic portions of a three-dimensional general relativistic
magnetohydrodynamic solver that is being developed by the authors.  We
studied the growth of the Papaloizou-Pringle instability in tori
orbiting Kerr black holes.  As was noted in H91, the parameter space
spanned by tori in the Schwarzschild spacetime is a rich one, and the
detailed behaviour of the unstable modes is very much model-dependent.  
This is certainly still true now that we have added black
role angular momentum to the mix.  However, several
general features do emerge from the study of tori in the Kerr spacetime.

Previous analytic and numerical work has established that the rate of
growth of unstable modes decreases as the width of the torus
increases.  Our results for models A3p, B3p, and B3r show that this
remains the case for tori orbiting Kerr black holes.

As in H91, we see a decrease in the density enhancement with torus width in the
bound tori.  However, only models A3p and B3p are bound tori, with $l > l_{mb}$,
(model B3r is sub-marginal, and the other three models are marginal).
Furthermore, as was noted in H91, the detailed behaviour of mode growth depends
sensitively on the choice of initial model, and mode growth dictates the
redistribution of matter and hence the magnitude of the density enhancements.

Two-dimensional studies have shown that accretion suppresses mode growth due to
the loss of the inner reflecting boundary, which is a prerequisite for mode
amplification.  This conclusion was supported by numerical and analytic work in
previous papers, as discussed in section 3.1.  However, the interplay between
accretion and mode growth in three-dimensional simulations is less clear.  The
argument has been made, in H91, that since inflow is primarily concentrated on
the equatorial plane, the loss of a reflecting boundary is localized to this
region, and mode growth can still occur given the right circumstances in
off-equator regions of the disk.  Models B3r, A3p, and B3p show that mode growth
in the absence of early inflow produces planets, while still generating
appreciable transient inflows as a consequence of the redistribution of matter
accompanying planet formation.  In the converse case, we have seen that early
inflow (marginal models E3r and X3p) effectively inhibits mode growth; the
establishement of a steady inflow coincides in these two models with the capping
of mode growth at a very low level, insufficient to disturb the disk.  Model E3p
would seem to straddle these two extremes in that it exhibits both early inflow
and the development of unstable modes.  
Model E3p, like models E3r and X3p, is a marginal torus.  E3p and X3p
are very similar in their initial conditions (see Table \ref{Param}),
although model E3p has an initial inner edge that is slightly farther
out that model X3p.  In spite of these structural similarities, model
E3p develops a significant PPI mode while E3r and X3p do not.  As
marginal tori, all of these three models show an early inflow of
matter.  However, the mass influx rate is less for model E3p (about
$-0.01$ in units normalized to the peak density) than it is for the
other marginal models (between $-0.04$ and $-0.08$).  Model E3p appears
to be a ``transition'' model, where the presence of (weak) early inflow
is insufficient to prevent mode growth.  Mode growth eventually begins,
and progresses much more slowly than in the other models that yield
planets; the growth rate of the $m=1$ mode for model E3p is one third
the rate of the next slowest mode growth, for model B3p.  Blaes (1987)
found that although the PPI modes can be stabilized by accretion, there
is a finite transition that permits both accretion and mode growth.
E3p appears to be such a case.  These results for 3D tori in the Kerr
metric reinforce the idea that accretion through the inner boundary at
the equatorial plane suppresses the growth of the PPI, and that this
suppression is as effective in three dimensions as it is in two.

Perhaps most interestingly, we have seen evidence of frame-dragging in those
models that develop an accretion flow.  Models B3r and E3r show evidence of a
retrograde inflow changing its winding sense as it enters the innermost regions
near the static limit, and model B3p shows a prograde accretion flow getting
smeared into a ring in the vicinity of the static limit.  As has been suspected
from the earliest days of numerical simulations of GR hydrodynamics, it is in
the deepest regions of the black hole potential well that we find the most
interesting, and perhaps most astrophysically distinct consequences of general
relativity.

\acknowledgements{This work was supported by NSF grant AST-0070979 and
NASA grant NAG5-9266.  The simulations were carried out on the Origin
2000 system at NCSA, and the Bluehorizon system of NPACI.}



\begin{thebibliography}

\bibitem[Balbus \& Hawley (1991)]{BH:91} Balbus, S.~A., \& Hawley, J.
~F. 1991, ApJ, 376, 214
\bibitem[Blaes (1987)]{B:87} Blaes, O. 1987, MNRAS, 227, 975
\bibitem[Blaes \& Hawley (1988)]{BH:88} Blaes, O., \& Hawley, J.~F.
1988, ApJ, 326, 277
\bibitem[Dwarkadas, \& Balbus (1996)]{db96} Dwarkadas, V., \& Balbus,
S.~A. 1996, ApJ, 467, 87
\bibitem[Elvis, Risaliti, \& Zamorani 2002]{ERZ:02} 
Elvis, M., Risaliti, G., \&  Zamorani, G. 2002, ApJ, 565, L75
\bibitem[Frolov, \& Novikov (1998)]{FN:98} Frolov, V.~P., \& Novikov, I.~D. 
1998, Black Hole Physics (Dordrecht:  Kluwer Academic)
\bibitem[Gat, \& Livio (1992)]{GL:92} Gat, O. \& Livio, M. 1992, ApJ,
396, 542
\bibitem[Hawley (1991)]{JFH:91} Hawley, J.~F. 1991, ApJ, 381, 496 (H91)
\bibitem[Hawley (1987)]{JFH:87} Hawley, J.~F. 1987, MNRAS, 225, 677
\bibitem[Hawley (1986)]{jfh86} Hawley, J.~F. 1986, in Radiation
Hydrodynamics in Stars and Compact Objects, ed. D. Mihalas and 
K.-H. Winkler (New York: Springer-Verlag), 369
\bibitem[Hawley \& Balbus (2002)]{hb02} Hawley, J.~F., \& Balbus,
S.~A. 2002, ApJ, submitted
\bibitem[Hawley, \& Smarr (1986)]{hs86} Hawley, J.~F., \& Smarr, L.~L.
1986, in Magnetospheric Phenomena in Astrophysics, ed. R. Epstein and W.
Feldman (New York: AIP), 263
\bibitem[Hawley, Smarr \& Wilson (1984a)]{HSW1:84} 
Hawley, J.~F., Smarr, L.~L., \& Wilson, J.~R., 1984a, ApJ, 277, 296 (HSWa)
\bibitem[Hawley, Smarr \& Wilson (1984b)]{HSW2:84} 
Hawley, J.~F., Smarr, L.~L., \& Wilson, J.~R., 1984b, ApJS, 55, 211 (HSWb)
\bibitem[Igumenshchev \& Beloborodov 1997]{igu97}Igumenshchev, I. V. \&
 Beloborodov, A. M. 1997, MNRAS, 284, 767
\bibitem[Koide, Shibata, \& Kudoh (1999)]{ksk99} Koide, S., Shibata, K., 
\& Kudoh, T. 1999, ApJ, 522, 727
\bibitem[Komissarov (1999)]{kom99} Komissarov, S. S. 1999, MNRAS, 303, 343
\bibitem[Lynden-Bell 1969]{lb69} Lynden-Bell, D. 1969, Nature, 223, 690
\bibitem[Misner, Thorne, \& WHeeler [1973)]{MTW} Misner, C.~W., Thorne,
K.~S., \& Wheeler, J.~ A, 1973, Gravitation (San Francisco: W.H. Freeman)
\bibitem[Narayan, Goldreich, \& Goodman (1987)]{NGG:87} 
Narayan, R., Goldreich, P., \& Goodman, J. 1987, MNRAS, 228, 1
\bibitem[Papaloizou \& Pringle (1984)]{PP:84} 
Papaloizou, J.~C.~B., \& Pringle, J.~E. 1984, MNRAS, 208, 721
\bibitem[Tanaka, et al. (1995)]{tan95} Tanaka, Y., et al. 1995, Nature, 
375, 659
\bibitem[Tremaine 1997]{tre97} Tremaine, S. 1997, in Unsolved Problems
in Astrophysics, eds.  J.N. Bahcall and J.P. Ostriker, (Princeton:
Princeton University), 137
\bibitem[van Paradijs \& McClintock (1995)]{vp95} van Paradijs, J., \&
McClintock, J.~E. 1995, in X-ray Binaries, eds.  W.H.G. Lewin, J. van
Paradijs, and E.P.J. van den Heuvel (Cambridge:  Cambridge University), 58
\bibitem[Wilson (1972)]{w72} Wilson, J. R. 1972, ApJ, 173, 431
\bibitem[Wilson (1975)]{w75} Wilson, J. R. 1975, Ann.~N.Y.~Acad.~Sci., 
262, 123
\bibitem[Wilson (1978)]{w78} Wilson, J. R. 1978, Proceedings of
International School of Physics Fermi Course LXV, ed. R. Giacooni and
R. Ruffini (Amsterdam: North Holland), 644
\bibitem[Yokosawa (1995)]{yok95} Yokosawa, M. 1995, PASJ, 47, 605

\end{thebibliography}
\end{document}